\documentclass[ twocolumn,superscriptaddress, amsmath, showpacs,
tightenlines,twocolumn,twoside,pra]{revtex4-2}%
\usepackage{amsfonts}
\usepackage{amstext}
\usepackage{amsmath}
\usepackage{amssymb}
\usepackage{latexsym}
\usepackage{graphicx,epstopdf}
\usepackage{ulem}
\usepackage[colorlinks,linkcolor=blue,anchorcolor=blue,citecolor=blue]%
{hyperref}
\usepackage{times}
\usepackage{subfigure}
\usepackage{color}
\usepackage{dcolumn}
\usepackage{graphicx}
\usepackage{soul}
\usepackage[dvipsnames, svgnames, x11names]{xcolor}
\usepackage{framed}
\usepackage{easyReview}%
\usepackage{colortbl}
\providecommand{\U}[1]{\protect\rule{.1in}{.1in}}

\newcommand{\CV}[0]{\color{black}}
\newcommand{\CIV}[0]{\color{black}}

\begin{document}
\title{Two-dimensional topological effect in a transmon qubit array with tunable couplings}
\author{Yan-Jun Zhao}
\thanks{The authors contributed equally to the work.}
\email{zhao\_yanjun@bjut.edu.cn}
\affiliation{Key Laboratory of Opto-electronic Technology, Ministry of Education, Beijing 	University of Technology, Beijing 100124, China}
\author{Yu-Qi Wang}
\thanks{The authors contributed equally to the work.}
\affiliation{Key Laboratory of Opto-electronic Technology, Ministry of Education, Beijing	University of Technology, Beijing 100124, China}
\author{Yang Xue}
\thanks{The authors contributed equally to the work.}
\affiliation{School of integrated circuits, Tsinghua University, Beijing 100084, China}
\affiliation{Frontier Science Center for Quantum Information, Beijing 100084, China}
\author{Xun-Wei Xu }
\affiliation{Key Laboratory of Low-Dimensional Quantum Structures and Quantum Control of	Ministry of Education, Department of Physics and Synergetic Innovation Center for Quantum Effects and Applications, Hunan Normal University, Changsha	410081, China}
\author{Yan-Yang Zhang}
\affiliation{School of Physics and Materials Science, Guangzhou University, 510006 Guangzhou, China}
\author{Wu-Ming Liu}
\affiliation{Beijing National Laboratory for Condensed Matter Physics, Institute of Physics, Chinese Academy of Sciences, Beijing 100190, China}
\affiliation{School of Physical Sciences, University of Chinese Academy of Sciences, Beijing 100190, China}
\author{Yu-xi Liu}
\email{yuxiliu@mail.tsinghua.edu.cn}
\affiliation{School of integrated circuits, Tsinghua University, Beijing 100084, China}
\affiliation{Frontier Science Center for Quantum Information, Beijing 100084, China}

\begin{abstract}
We investigate a square-lattice architecture of superconducting transmon
qubits with inter-qubit interactions mediated by inductive couplers. Via
periodically modulating the couplers, the Abelian gauge potential, termed
effective magnetic flux, can be synthesized artificially, making the system an
excellent platform for simulating two-dimensional topological physics.
\CV
First, we focus on the three-leg ladder which only has three rows and
investigate the chiral dynamics therein for the single-particle ground state
when the effective magnetic flux varies. We find what we call the
\textquotedblleft staggered vortex-Meissner phase transition\textquotedblright%
, where the vortex number can typically stagger a few times between one
(defined as \textquotedblleft vortex phase\textquotedblright) and larger integers
(defined as \textquotedblleft Meissner phase\textquotedblright) when the effective magnetic 
flux changes between $-\pi$ and $\pi$. This phenomenon, actually not a phase transition by definition, is quite different from the vortex-Meissner phase transition in the two-leg ladder that, in
contrast, possesses only two rows and is usually treated as the
quasi-two-dimensional model. Also, we find that the
chiral current relies on the effective magnetic flux according to a squeezed
sinusoidal function. Both the staggering of the vortex number and squeezing of the chiral current
can be controllable by the coupling ratio, which is defined by the coupling-strength ratio between the column direction and row direction. Second,
we continue to increase the number of rows beyond three, and the topological band
structure to be anticipated at an infinite number of rows begins to occur even
for a relatively small number (ten or so for typical parameters) of rows.
\CIV
This heralds a small circuit scale
\CV
enough
\CIV
to observe the topological band. The behavior of the edge state in the band
gap can be interpreted by the topological Chern number%
\CV
, which
\CIV
can be calculated through integrating the Berry curvature with respect to the
first Brillouin zone.
\CV
Last%
\CIV
, we present a systematic method on how to measure the topological band
structure based on time- and space-domain Fourier transformation of the wave
function after properly
\CV
exciting the qubits, which should be helpful for comprehensively analyzing the
topological physics since all the topological properties are mainly contained in
the band structure. Our results offer
\CIV
an avenue for simulating two-dimensional topological physics on the
state-of-the-art superconducting quantum chips.
  
\end{abstract}
\maketitle

\section{Introduction}

In the recent years, a few pioneering works have emerged in superconducting
quantum circuits%
~%
\cite{Krantz2019APR} that focused on quantum error correction%
~%
\cite{Barends2014Nature,Chow2014NC,Gong2022NSR}, quantum
\CV
supremacy
\CIV
demonstration%
~%
\cite{Arute2019Nature,Wu2021PRL,ZhuQingLing2022SB}, and even quantum chemistry
simulation%
~%
\cite{OMalley2016PRX,Kandala2017Nature,Kandala2020Science}. Fundamentally,
these significant achievements can be attributed to the upgrade of the
integration level and lifetime of superconducting qubits. For example, the
recent Zuchongzhi processor%
~%
\cite{Wu2021PRL} has achieved as high as 66 functional qubits with a mean
lifetime 30.6%
~%
microseconds in an architecture of tunable couplings and tunable frequencies.
Besides, the
\CV
more recent
\CIV
Eagle processor%
~%
\cite{Chow2021-IBM} has reached to date the maximum qubit number 117 in a
different architecture from Zuchongzhi. On the other hand, adopting the
tantalum material, the qubit has elevated the lifetime to the remarkable
hundreds of microseconds%
~%
\cite{Place2021NC,Wang2022npj}. The state-of-the-art integration level and
lifetime heralds the noisy intermediate-scale
\CV
quantum
\CIV
(NISQ) era%
~%
\cite{Preskill2018Quantum,Kjaergaard2020ARCMP}.

One typical application of near-term NISQ devices is known as condensed matter
physics simulation. On this topic, there have been extensive proposals and
experiments based on single or several qubits%
~%
\cite{Leak2007Science,Berger2012PRB,Berger2013PRA,Schroer2014PRL,Zhang2017PRA,Roushan2014Nature,Flurin2017PRX,Ramasesh2017PRL,Tan2017npj,Tan2018PRL,Zhong2016PRL,Guo2019PRapp}%
, and also multiple qubits%
~%
\cite{Roushan2017NP,Nunnenkamp2011NJP,Mei2015PRA,Yang2016PRA,Tangpanitanon2016PRL,Gu2017arXiv,Roushan2017Science,Xu2018PRL,Yan2019Science,Ye2019PRL,Zhao2020PRA,Guan2020PRA,Arute2019Nature,Ge2021PRB,Mi2021Nature}%
. However, most multiple-qubit studies focused on one-dimension chains
~%
\cite{Nie2021PRL,Zhang2022PRapplied,Roushan2017NP,Roushan2017Science,Xu2018PRL,Yan2019Science,Mi2021Nature}
or quasi-two-dimensional \CV two-leg \CIV ladders%
~%
\cite{Ye2019PRL,Zhao2020PRA,Guan2020PRA}, and true-two-dimensional lattices%
~%
\cite{Yang2016PRA,Ge2021PRB} are less studied. Indeed, the integration level
and lifetime of the NISQ device make it an excellent platform to conduct
two-dimensional simulation of condensed matter physics. In particular, the
multiple-qubit quantum behavior therein is an appealing topic which merits
further investigation. Thus, it makes great sense to study two-dimensional
simulation of condensed matter physics.

As is well
\CV
known%
\CIV
, the quantum Hall effect is a renowned phenomenon in two-dimensional
condensed matter physics. Characterized by the Harper Hamiltonian, this
phenomenon features the electron moving in a square lattice penetrated by
uniform magnetic fields%
~%
\cite{Hatsugai1993PRL,Hatsugai1993PRB}. As has been shown, for neutral atoms
in an optical lattice, the Harper Hamiltonian can be similarly synthesized
with the artificial magnetic fields, which can be engineered via periodically
modulating the onsite energy%
~%
\cite{Aidelsburger2011PRL,Aidelsburger2013PRL,Miyake2013PRL} or classically
driving the atomic internal states%
~%
\cite{Celi2014PRL}. Compared to cold atoms, superconducting qubits possess the
convenience of tunability and scalability. Besides, there is already the
experimental implementation of a \textquotedblleft
one-dimensional\textquotedblright\ Harper Hamiltonian using interacted
transmon qubits mediated by inductive couplers%
~%
\cite{Roushan2017Science}. However, how to synthesize the Harper Hamiltonian
using two-dimensional superconducting NISQ
\CV
circuits
\CIV
needs to be exhaustively studied.

Meanwhile, we note that the artificial magnetic fields can be synthesized via
periodically modulating the inductive couplers in a triangle unit of transmon
qubits
~%
\cite{Roushan2017NP}. This inspires us to further apply the inductive couplers
to a square array of transmon qubits and further engineer the Harper
Hamiltonian.
\CV
We just note that very recently, the two-dimensional fractional quantum Hall
effect has been experimentally studied by periodically modulating another type
of transmon-based couplers%
~%
\cite{Wang2024Science}.
\CIV
Although it has been claimed that the inductive couplers can be applied to
demonstrate the fractional quantum Hall effect, where the triangle and
square-lattice models are discussed, no concrete circuits that realize these
models have therein be calculated in detail%
~%
\cite{Ge2021PRB}. Simultaneously, the inductive coupler can perfectly switch
off the inter-qubit couplings, thus effectively avoiding the problem of
frequency crowding%
~%
\cite{Chen2014PRL}. However, the similar feature is not discussed in other
artificial-magnetic-field-synthesizing schemes based on the nonlinearity of
Josephson junctions%
~%
\cite{Koch2010PRA,Yang2016PRA}, and not possessed in those based on fixed
couplings with periodically-modulated onsite energy%
~%
\cite{Alaeian2019PRA,Zhao2020PRA}. \CV Besides improving the gate fidelity in quantum computation, the inductive coupler can also benefit the precise generation and measurement of qubit sates in quantum simulation, where, sometimes, we also need to alternatively decouple the interaction between particular qubits. \CIV

\CV
Although the multiple-particle state describes a real material better, which is typically a fermionic system (e.g., electron gas in potentials at zero temperature that occupy all the energy states below the Fermi level), here we prefer to focus on the the single-particle state, just originating from two particular reasons. First, 
we have already noted that in the three-leg ladder (two-dimensional square lattice model with just three rows), multiple fermionic cold atoms below the Fermi level can occupy both the edge states and the bulk states, and thus exhibit interesting
chiral dynamics~\cite{Celi2014PRL,Mancini2015Science}. However, it remains to be investigated what the chiral dynamics is like in the single-particle
case. Second, the theoretical model of transmon qubits
are two-level systems, which can not be directly mapped to fermions. Luckily, by virtue of Jordan-Wigner transformation, the one-dimensional chain of coupled qubits have been successfully bridged to a practical fermionic system%
~%
\cite{Mei2013PRB}.
However, this transformation proves ineffective to the approach two-dimensional qubit array, because it will induce extra operator-dependent phase factors to the effective coupling strengths and thus makes the final Hamiltonian far from a real fermionic system~\cite{Derzhko2001JPS}.
Exclusively, such extra phase factors will vanish for the special single-particle state, implying the Pauli raising (lower) operators can be directly mapped to the fermionic creator (annihilation) operators. Thus, in contrast to the multiple-particle scenario, the
single-particle-ground-state chiral dynamics is worthy of prior investigation, in the two-dimensional qubit array with effective magnetic flux we will work on here.
\CIV

In this paper, we propose to engineer the Harper Hamiltonian in a
two-dimensional architecture based on interacting transmon qubits mediated by
inductive couplers. To be concrete, we first investigate the \CV three-leg ladder
model with effective magnetic flux \CIV (i.e., three-row Harper Hamiltonian), and find a novel \CV\textquotedblleft staggered vortex-Meissner
phase transition\textquotedblright\, which, not an actual phase transition by definition, is quite \CIV different from the \CV vortex-Meissner phase transition \CIV
in the two-leg ladder model. This \CV so-called ``phase transition'' results from the finite size of the lattice length and can be controlled by \CIV the
competition between the coupling strengths and magnetic flux. Then, we study
the variation of the topological band structure when the row number is
increased. In this way, feasible qubit numbers can be suggested for simulating
observable topological phenomena \CV with currently-available technology\CIV. \CV The possibility to observe Hofstadter-butterfly spectrum will also be discussed. For the phenomena studied, we will also propose the experimental measurement method based on superconducting quantum circuits.\CIV

In Sec.%
~%
\ref{sec:Model}, we introduce the transmon architecture with
inductive-coupler-mediated interactions, from which the Harper Hamiltonian can
be further derived. In Sec.%
~%
\ref{sec:phase transition}, we \CV study \CIV the vortex number and chiral current in
the double-ladder model for different magnetic fluxes and coupling strengths.
\CV In Sec.~\ref{sec:multi-band model}\CIV, 
we \CV investigate \CIV the topological effect when the row number is
increased beyond three. 
\CV In  Sec.~\ref{sec:HotBut}, we analyze the Hofstadter butterfly spectrum in the proposed two-dimensional transmon array with artificial magnetic flux. \CIV
 In Sec.%
~%
\ref{Sec:ExpDetails}, we discuss the experiment details on how to generate the
single-particle ground state and measure the chiral currents and topological
energy bands. In Sec.%
~%
\ref{sec:DC}, we summarize the results. \ 

\section{Qubit architecture with inductive couplers \ }

\label{sec:Model}

\subsection{Circuit model}%

\begin{figure*}[ptb]
\centering\includegraphics[
width=1\textwidth,clip
]{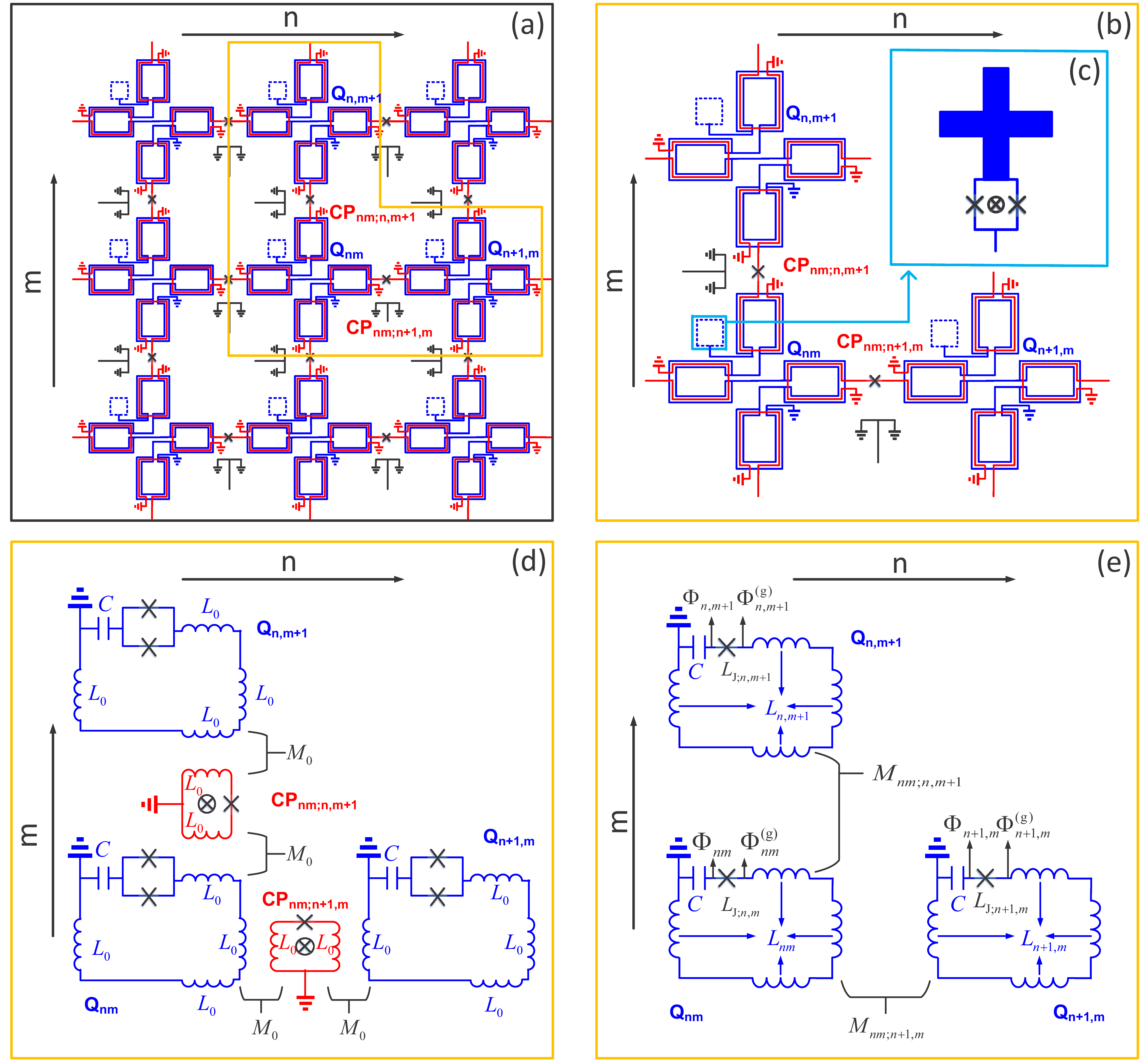}
\caption{(a) In the two-dimensional square lattice of transmons,
the qubit at the site $nm$ (denoted by $\mathrm{Q}_{nm}$ and colored blue)
interacts with its nearest neighbours with
an inductive coupler (colored red).
Here, the horizontal and vertical coordinates, $n$ and $m$,
represent the column and row indices.
The dashed blue squares denote the main part of qubit [see (c) for concrete structure].
(b) The coupling details between the qubit at $nm$ ($\mathrm{Q}_{nm}%
$) and its right ($\mathrm{Q}_{n+1,m}$) and upper ($\mathrm{Q}_{n,m+1}%
$) nearest neighbours. The transmon qubit is grounded via a specially meandering wire
that acts as a gradiometer to cancel the environmental magnetic flux noise, and through which,
the qubit inductively couples to each adjacent coupler (e.g., $\mathrm
{CP}_{nm;n+1,m}$ or $\mathrm{CP}_{nm;n,m+1}$).
The coupler consists of a Josephson junction and is tunable via the bias magnetic flux.
(c) Main part of the qubit: the SQUID constituting the transmon connects to the cross-shaped capacitor that grounds the transmon
that couples to the driving
and read-out signals~\cite{Barends2013PRL}.
(d) Equivalent circuit diagram of the coupling mechanism shown in (c).
The SQUID is shunted via a self capacitance ($C$)
and located in a loop with four segments of identical self inductances ($L_0$).
Each segment simultaneously couples to the adjacent coupler with a mutual inductance $M_0$.
And the coupling segment in the coupler also possesses a self inductance $L_0$.
The coupler junction inductance is $L_\mathrm{T}$.
(e) Simplified circuit diagram from (d), where the SQUID (superconducting quantum interference device)
is modelled as one single junction (with the \CV equivalent \CIV junction inductance \CV$L_{\mathrm{J};nm}%
$ \CIV for the site $nm$, which is tunable by external magnetic flux threaded the SQUID loop),
and the coupler-mediated indirect qubit-qubit coupling is replaced
by a tunable mutual inductive coupling (e.g., with the mutual inductance
$M_{nm;n+1,m}$ between the site $nm$ and $n,m+1$).
The resulting total inductance around the qubit loop is $L_{nm}%
$ at the site $nm$. See Appendix.~\ref{Append:Int-coupler} for details.
To construct the Hamiltonian,
the node fluxes $\Phi_{nm}$ and $\Phi_{nm}^{(\mathrm{g})}%
$ are chosen respectively at the two terminals of the qubit junction at the site $nm$.}
\label{fig:model}
\end{figure*}%

We investigate a square array of transmon qubits with inter-qubit interactions
mediated by inductive couplers. As schematically shown in Fig.%
~%
\ref{fig:model}(a), each qubit (colored blue) at the site $nm$ (concise
abbreviation for $n,m$ without causing any ambiguity) couples intermediately
via the coupler (colored red) to its four nearest neighbours at sites
$n^{\prime}m^{\prime}=n\pm1,m$ and $n,m\pm1$. In detail, the qubit is grounded
via a wound wire that consists of four different inductive segments [see
Figs.~\ref{fig:model}(b)], which can operate as an gradiometer aiming at
eliminating the homogeneous electromagnetic noise. The wire is plugged out
from the SQUID (superconducting quantum interference device) that constitutes
the transmon qubit [see Fig.%
~%
\ref{fig:model}(c)]. The qubit frequency can be controlled through the
magnetic flux piercing the SQUID and the excitation and measurement of the
qubit are realized by the cross-shaped capacitor that couples to the driving
field and readout resonator ~\cite{Barends2013PRL}. Likewise, the inductive
loop of the coupler [see Fig.%
~%
\ref{fig:model}(d)] is also designed as a gradiometer to cancel the
homogeneous electromagnetic noise. The tunability of the coupler is guaranteed
by the externally applied magnetic flux.

To be more intuitive, we now simplify the concrete circuit in Fig.%
~%
\ref{fig:model}(a) into a more general schematic in Fig.%
~%
\ref{fig:model}(e). Because of the externally applied magnetic flux
$\Phi_{nm;n^{\prime}m^{\prime}}$ , the coupler junction between any qubit site
$nm$ and its nearest neighbor $n^{\prime}m^{\prime}$ can be identified with a
linear inductor to the small quantum signal%
~%
\cite{Geller2015PRA}. This implies that the indirect qubit interaction
mediated by the coupler can be modelled by a linear inductor network that
describes the interplay between the SQUID branch currents $I_{nm}$ and
$I_{n^{\prime}m^{\prime}}$ at $nm$ and $n^{\prime}m^{\prime}$. Therein, the
mutual inductance possesses the form \CV(see Appendix.%
~%
\ref{Append:Int-coupler})\CIV
\begin{equation}
M_{nm;n^{\prime}m^{\prime}}=-\frac{M_{0}^{2}\cos\left(  \frac{2\pi}{\Phi_{0}%
}\Phi_{nm;n^{\prime}m^{\prime}}\right)  }{L_{\text{T}}+2L_{0}\cos\left(
\frac{2\pi}{\Phi_{0}}\Phi_{nm;n^{\prime}m^{\prime}}\right)  },\label{eq:M}%
\end{equation}
and the self inductance in series is $L_{nm}=4L_{0}+\sum_{n^{\prime}m^{\prime
}\in%
\mathbb{C}
_{nm}}M_{nm;n^{\prime}m^{\prime}}$ at the site $nm$ . Here, the symbol $%
\mathbb{C}
_{nm}$ represents all the four nearest neighbours of the site $nm$, $\Phi_{0}
$ the magnetic flux quantum, $L_{\text{T}}=\Phi_{0}/2\pi I_{\text{c}}$ the
junction inductance of the coupler, $I_{\text{c}}$ the critical current of the
coupler junction, $M_{0}$ the mutual inductance between the coupler loop and
the qubit loop grounding the SQUID, $\ $and $L_{0}$ the self inductance of one
inductive segment of the gradiometer-shaped coupling loop. We stress that the
two-site symbol is invariant if we exchange the order of the site indices,
e.g., $M_{nm;n^{\prime}m^{\prime}}=M_{nm;n^{\prime}m^{\prime}}$.

To perform the canonical quantization%
~%
\cite{Devoret1995Les}, we use the node flux $\Phi_{nm}$ ($\Phi_{nm}^{\left(
\mathrm{g}\right)  }$) to represent the node flux between the SQUID and
shunting capacitor (gradiometer-shaped grounding wire) at the site $nm$. Then,
we can give the Lagrangian of the whole circuit, that is,
\begin{align}
L=  &  \sum_{nm}\frac{C\dot{\Phi}_{nm}^{2}}{2}+E_{\text{J,}nm}\cos\left[
\frac{2\pi}{\Phi_{0}}\left(  \Phi_{nm}-\Phi_{nm}^{\left(  \mathrm{g}\right)
}\right)  \right] \nonumber\\
&  -\sum_{nm}%
\CV
\left(  \frac{L_{nm}I_{nm}^{2}}{2}+M_{nm;n+1,m}I_{nm}I_{n+1,m}\right)
\CIV
\nonumber\\
&  -\sum_{nm}M_{nm;n,m+1}I_{nm}I_{n,m+1},
\end{align}
where the two terms in the first line respectively denote the capacitive and
Josephson inductive energies, while the last two lines represent the inductive
energy induced by the coupler and surrounding circuit. Besides, the magnetic
flux $\Phi_{nm}^{(\mathrm{g})}$, which originates from the currents flowing
through the SQUID at $nm$ and all its four nearest neighbours, takes the form
\begin{equation}
\Phi_{nm}^{(\mathrm{g})}=L_{nm}I_{nm}+\sum_{n^{\prime}m^{\prime}\in%
\mathbb{C}
_{nm}}M_{nm;n^{\prime}m^{\prime}}I_{n^{\prime}m^{\prime}},\label{eq:Phi_g}%
\end{equation}
where $%
\mathbb{C}
_{nm}$ means all the four nearest neighbours of the site $nm$ as mentioned%
\CV
, and the notation $I_{nm}$ means the currents flowing through the SQUIDs at
$nm$%
\CIV
. Additionally, $C$ is the homogeneous transmon shunting capacitance,
\CV
$E_{\mathrm{J,}nm}=\Phi_{0}^{2}/4\pi^{2}L_{\text{J},nm}=\left(  \Phi_{0}%
/2\pi\right)  I_{\text{c};nm}$
\CIV
the Josephson energy, and $L_{\text{J},nm}=\frac{\Phi_{0}}{2\pi I_{\text{c}%
;nm}}$ the equivalent junction inductance of the SQUID, which is regarded as
one single junction [see Fig.~\ref{fig:model}(e)]. According to the canonical
quantization, the node charge takes $Q_{nm}=\partial L/\partial\dot{\Phi}%
_{nm}$, and the full Hamiltonian can be given by $H=\sum_{nm}Q_{nm}\dot{\Phi
}_{nm}-L$, yielding%
\begin{align}
H=  &  \sum_{nm}\left\{  \frac{Q_{nm}^{2}}{2C}-E_{\text{J},nm}\cos\left[
\frac{2\pi}{\Phi_{0}}\left(  \Phi_{nm}-\Phi_{nm}^{\left(  \mathrm{g}\right)
}\right)  \right]  \right\} \nonumber\\
&  +\sum_{nm}%
\CV
\left(  \frac{L_{nm}I_{nm}^{2}}{2}+M_{nm;n+1,m}I_{nm}I_{n+1,m}\right)
\CIV
\nonumber\\
&  +\sum_{nm}M_{nm;n,m+1}I_{nm}I_{n,m+1}.\label{eq:H}%
\end{align}%
\CV
Here, we assume the coupling is sufficiently weak, which means $\left\vert
M_{nm;n^{\prime}m^{\prime}}\right\vert ,L_{nm}\ll L_{\text{J};nm}$ for
$n^{\prime}m^{\prime}$ taking the nearest neighbours of $nm$ [see Fig.%
~%
\ref{fig:model}%
(e)]. Thus the flux drop across the junction is dominant, that is, $\left\vert
\Phi_{nm}^{\left(  \mathrm{g}\right)  }\right\vert \ll\left\vert \Phi
_{nm}\right\vert $ and then the cosine function in Eq.%
~%
(\ref{eq:H}) can be approximated by the first-order expansion with respect to
$\Phi_{nm}^{\left(  \mathrm{g}\right)  }$, giving
\begin{align}
&  E_{\text{J},nm}\cos\left[  \frac{2\pi}{\Phi_{0}}\left(  \Phi_{nm}-\Phi
_{nm}^{\left(  \mathrm{g}\right)  }\right)  \right] \nonumber\\
&  \approx E_{\text{J},nm}\cos\left(  \frac{2\pi}{\Phi_{0}}\Phi_{nm}\right)
+E_{\text{J},nm}\sin\left(  \frac{2\pi}{\Phi_{0}}\Phi_{nm}\right)  \frac{2\pi
}{\Phi_{0}}\Phi_{nm}^{\left(  \mathrm{g}\right)  }\nonumber\\
&  \approx E_{\text{J},nm}\cos\left(  \frac{2\pi}{\Phi_{0}}\Phi_{nm}\right)
+I_{nm}\Phi_{nm}^{\left(  \mathrm{g}\right)  },\label{eq:Ecos}%
\end{align}
where, also due to $\left\vert \Phi_{nm}^{\left(  \mathrm{g}\right)
}\right\vert \ll\left\vert \Phi_{nm}\right\vert $, the explicit approximation
$I_{nm}=I_{\text{c},nm}\sin\left[  \frac{2\pi}{\Phi_{0}}\left(  \Phi_{nm}%
-\Phi_{nm}^{\left(  \mathrm{g}\right)  }\right)  \right]  \approx
$\ $I_{\text{c},nm}\sin\left(  \frac{2\pi}{\Phi_{0}}\Phi_{nm}\right)  $ has
been made. From the expression of $\Phi_{nm}^{\left(  \mathrm{g}\right)  }$ in
Eq.%
~%
(\ref{eq:Phi_g}), we must point out that when the index $nm$ in Eq.%
~%
(\ref{eq:H}) sweeps all the sites, both the terms $I_{nm}\Phi_{nm}^{\left(
\mathrm{g}\right)  }$ and $I_{n+1,m}\Phi_{n+1,m}^{\left(  \mathrm{g}\right)
}$ ($I_{nm}\Phi_{nm}^{\left(  \mathrm{g}\right)  }$ and $I_{n,m+1}\Phi
_{n,m+1}^{\left(  \mathrm{g}\right)  }$) will contribute to the mutual
inductive energy $M_{nm;n+1,m}I_{nm}I_{n+1,m}$ ($M_{n,m+1;n,m}I_{nm}I_{n,m+1}
$). To this end,
\CIV
the full Hamiltonian in Eq.%
~%
(\ref{eq:H}) can now be transformed into%
\begin{equation}
H=\sum_{nm}H_{nm}+U_{nm},
\end{equation}
where $H_{nm}$ is the free qubit Hamiltonian at the site $nm$, and $U_{nm}$
the interaction Hamiltonian between the site $nm$ and its right and upper
nearest neighbours, that is,%
\begin{align}
H_{nm}=  &  \frac{Q_{nm}^{2}}{2C}-E_{\text{J},nm}\cos\left(  \frac{2\pi}%
{\Phi_{0}}\Phi_{nm}\right)  -\frac{L_{nm}I_{nm}^{2}}{2},\label{eq:Hnm_free}\\
U_{nm}=%
\!%
&  -%
\!%
M_{nm;n+1,m}I_{nm}I_{n+1,m}%
\!%
-%
\!%
M_{nm;n,m+1}I_{nm}I_{n,m+1}.\label{eq:Unm}%
\end{align}

In the transmon regime%
~%
\cite{Koch2007PRA}, the Josephson energy $E_{\text{J},nm}$ is much larger than
the charging energy $E_{\text{C}}=e^{2}/2C$ (e.g., $E_{\text{J},nm}/$
$E_{\text{C}}\sim50$, and $e$ being the elementary charge) characterizing the
capacitive energy. And then the qubit circuit mimics a virtual particle well
localized in the vicinity of the Josephson potential energy bottom:
$\left\vert \Phi_{nm}\right\vert \ll\Phi_{0}$. Thus, we can expand the free
qubit Hamiltonian to the quartic order, giving
\begin{equation}
H_{nm}=\frac{Q_{nm}^{2}}{2C}+\frac{\Phi_{nm}^{2}}{2L_{\text{J},nm}}-\frac
{1}{24}\frac{1}{L_{\text{J},nm}}\left(  \frac{2\pi}{\Phi_{0}}\right)  ^{2}%
\Phi_{nm}^{4},\label{eq:Hnm}%
\end{equation}
Here, the term $\frac{1}{2}L_{nm}I_{nm}^{2}$ has been neglected by the
assumption $L_{nm}\ll L_{\text{J},nm}$.\ 

\subsection{Derivation of the Harper model}

Now, we represent the node flux $\Phi_{nm}$ and node charge $Q_{nm}$ with
bosonic annihilation and creation operators $a_{nm}$ and $a_{nm}^{\dag}$,
i.e.,
\begin{align}
\Phi_{nm}  &  =\sqrt{\frac{\hbar Z_{nm}}{2}}\left(  a_{nm}+a_{nm}^{\dag
}\right)  ,\label{eq:Phinm}\\
Q_{nm}  &  =\sqrt{\frac{\hbar}{2Z_{nm}}}\frac{\left(  a_{nm}-a_{nm}^{\dag
}\right)  }{i}.\label{eq:Qnm}%
\end{align}
Here, the parameter $Z_{nm}=\sqrt{L_{\text{J},nm}/C}$ is called as the qubit
impedance. Substituting Eqs.%
~%
(\ref{eq:Phinm}) and (\ref{eq:Qnm}) into Eq.%
~%
(\ref{eq:Hnm}), the free Hamiltonian is transformed into%
\begin{equation}
H_{nm}=\hbar\omega_{nm}a_{nm}^{\dag}a_{nm}-\frac{E_{\text{C}}}{2}a_{nm}%
^{\dag2}a_{nm}^{2},\label{eq:Hq}%
\end{equation}
where $\omega_{nm}=\omega_{\text{p,}nm}-E_{\text{C}}/\hbar$ is the qubit
frequency, $\omega_{\text{p,}nm}=1/\sqrt{CL_{\text{J},nm}}=\sqrt{8E_{\text{C}%
}E_{\text{J,}nm}}/\hbar$ the Josephson plasma frequency, and only the
number-conserving terms are kept. Due tot the nonlinearity $E_{\text{C}}$, the
the transmon can also be represented as a two-level system.

Hereafter, we will mainly focus on the case of the single-particle excitation,
which means the nonlinear term $a_{nm}^{\dag2}a_{nm}^{2}$ can be neglected in
the free Hamiltonian, thus making Eq.~(\ref{eq:Hq}) become%
\begin{equation}
H_{nm}=\hbar\omega_{nm}a_{nm}^{\dag}a_{nm}.
\end{equation}

When treating interaction Hamiltonian $U_{nm}$, we neglect the nonlinear
effect of the SQUID, thus simplifying the SQUID branch current into
$I_{nm}=\Phi_{nm}/L_{\text{J},nm}$. Based on this approximation, the
interaction Hamiltonian becomes
\begin{align}
U_{nm}=  &  \hbar G_{nm;n+1,m}a_{nm}\left(  a_{n+1,m}+a_{n+1,m}^{\dag}\right)
+\mathrm{H.c.}\nonumber\\
&  +\hbar G_{nm;n,m+1}a_{nm}\left(  a_{n,m+1}+a_{n,m+1}^{\dag}\right)
+\mathrm{H.c.,}\label{eq:V}%
\end{align}
where the first (second) line denotes the row (column) couplings, and the
coupling strength $G_{nm;n^{\prime}m^{\prime}}$ takes the expression%
\begin{equation}
G_{nm;n^{\prime}m^{\prime}}=-\frac{M_{nm;n^{\prime}m^{\prime}}}{2}\frac
{\omega_{\text{p,}nm}\omega_{\text{p,}n^{\prime}m^{\prime}}}{\sqrt
{Z_{nm}Z_{n^{\prime}m^{\prime}}}}.\label{eq:G}%
\end{equation}
\CV Here, note that $n'm'$ only takes the nearest neighbor of $nm$, i.e., $n^{\prime}m^{\prime}\in\mathbb{C}_{nm}$ .\CIV

As a first step to engineer effective magnetic flux in the Harper model%
~%
\cite{Hatsugai1993PRL,Hatsugai1993PRB}, the qubit frequencies should be
synthesized according to the parity of the row index $m$. To do this, we
assume that the SQUID's equivalent junction inductance takes $L_{\text{J}%
,nm}\equiv L_{\text{J},\text{o}}$ ($L_{\text{J},nm}\equiv L_{\text{J}%
,\text{e}}$) for the odd and even $m$, which can be conveniently achieved via
tuning the bias magnetic flux. This leads to a set of parity-dependent
parameters, which can be summarized as $\CV S\CIV_{nm}\equiv \CV S\CIV_{\text{o}}$
($\CV S\CIV_{nm}\equiv \CV S\CIV_{\text{e}}$) for the odd (even) $m$ where \CV the symbol $ S$  denotes the quantity \CIV
$\omega$ (qubit frequency), $\omega_{\text{p}}$ (plasma frequency), and $Z$
(qubit impedance). Then, the full free Hamiltonian can be rewritten as
\begin{align}
H_{0}  &  =\sum_{nm}H_{nm}\nonumber\\
&  =\sum_{\substack{n \\m~\text{odd}}}\hbar\omega_{\mathrm{o}}a_{nm}^{\dag
}a_{nm}+\sum_{\substack{n \\m~\text{even}}}\hbar\omega_{\mathrm{e}}%
a_{nm}^{\dag}a_{nm},
\end{align}

Secondly, the row coupling strengths are designed to be identical. For this
purpose, we also specify the static bias magnetic flux of the row coupler
according to the parity of the row index $m$, i.e., $\Phi_{nm;n+1,m}\equiv
\Phi_{\mathrm{o}}$ ($\Phi_{n,m;n+1,m}\equiv\Phi_{\mathrm{e}}$) for the odd
(even) row index $m$. This further results in parity-dependent row coupling
strengths, that is, $G_{nm;n+1,m}\equiv G_{\text{o}}$ ($G_{nm;n+1,m}\equiv
G_{\text{e}}$) for the odd (even) $m$. This, however, does not necessarily
mean $G_{\text{o}}\neq G_{\text{e}}$, since from Eqs.%
~%
(\ref{eq:M}) and (\ref{eq:G}), the difference caused by $\omega_{\text{o}}$
and $\omega_{\text{e}}$ as well as $Z_{\text{o}}$ and $Z_{\text{e}}$ can in
principle be compensated via properly tuning $\Phi_{\mathrm{o}}$ and
$\Phi_{\mathrm{e}}$. Hence, the homogeneous row coupling strengths can be
created as $G_{\text{o}}=G_{\text{e}}=-g_{x}$. After the rotating-wave
approximation is made for the intra-row coupling terms, the full interaction
Hamiltonian reduces to%
\begin{align}
U=  &  \sum_{nm}U_{nm}=-\sum_{nm}\hbar g_{x}a_{nm}a_{n+1,m}^{\dag
}+\mathrm{H.c.}\nonumber\\
&  +\sum_{nm}\hbar G_{nm;n,m+1}a_{nm}a_{n,m+1}^{\dag}+\mathrm{H.c..}%
\end{align}

The other critical step to synthesize the effective magnetic flux is biasing
each column coupler with both direct and alternating components, i.e.,%
\begin{equation}\label{eq:Phi_column}
\Phi_{nm;n,m+1}=\bar{\Phi}+\Phi_{\text{eff}}\cos\left( \CV \omega_{\mathrm{oe}}
\CIV t+\CV
\gamma'_{nm}\CIV\right)  .
\end{equation}
Here, $\CV\gamma_{nm}'\CIV\equiv-n\gamma$ ($n\gamma$) for odd (even) $m$, and
the modulation frequency matches the frequency difference between adjacent
rows, i.e., $\CV\omega_{\mathrm{oe}}\CIV=\omega_{\text{o}}-\omega_{\text{e}}>0$. Then, suppose
$L_{0}\ll L_{\text{T}}$, the column coupling strength should take the form
\begin{equation}
G_{nm;n,m+1}%
\!%
=%
\!%
2t_{y}\cos%
\!%
\left[  \frac{2\pi\bar{\Phi}}{\Phi_{0}}%
\!%
+%
\!%
\frac{2\pi\Phi_{\text{eff}}}{\Phi_{0}}\cos\left( \CV \omega_{\mathrm{oe}}\CIV t%
\!%
+%
\!%
\CV\gamma'_{nm}\CIV\right)  \right]  ,
\end{equation}
with $t_{y}=M_{0}^{2}\omega_{\text{p},\text{o}}\omega_{\text{p},\text{e}%
}/4L_{\text{T}}\sqrt{Z_{\text{o}}Z_{\text{e}}}$ denoting the bare column
coupling strength. Besides, we mention that $\omega_{\text{p},\text{o}}$
($\omega_{\text{p},\text{e}}$) is namely $\omega_{\text{p},nm}$ for the odd
(even) $m$ and simultaneously, $Z_{\text{p},\text{o}}$ ($Z_{\text{p},\text{e}%
}$) is $Z_{nm}$ for the odd (even) $m$.

Now, we discuss the inter-qubit couplings via entering the interaction picture
defined by $U_{0}=\exp\left(  -\frac{i}{\hbar}H_{0}t\right)  $. This gives the
time-dependent interaction Hamiltonian $H_{\text{I}}=U_{0}^{\dag}UU_{0}$%
,$\,$which takes the explicit form
\begin{align}
H_{\text{I}}=  &  -\sum_{nm}\hbar g_{x}a_{nm}a_{n+1,m}^{\dag}+\mathrm{H.c.}%
\nonumber\\
&  +\sum_{nm}\hbar G_{nm;n,m+1}Y_{nm}\left(  t\right)  +\mathrm{H.c.,}%
\end{align}
with the symbol $Y_{nm}\left(  t\right)  =a_{nm}a_{n,m+1}^{\dagger}e^{i\left(
\omega_{n,m+1}-\omega_{nm}\right)  t}$. Here, note that the column coupling
strength can also be expanded into the Fourier series as
\begin{align}
G_{nm;n,m+1}=  &  t_{y}\sum_{k=-\infty}^{\infty}i^{k}e^{ik\left(  \CV \omega_{\mathrm{oe}}\CIV t + \CV\gamma'_{nm}\CIV\right)  }J_{k}\left(  \frac{2\pi\Phi_{\text{eff}}}%
{\Phi_{0}}\right) \nonumber\\
&  \times\left[  e^{i\frac{2\pi\bar{\Phi}}{\Phi_{0}}}+\left(  -1\right)
^{k}e^{-i\frac{2\pi\bar{\Phi}}{\Phi_{0}}}\right]  ,
\end{align}
where $J_{k}\left(  x\right)  ~$stands for the $k$th Bessel function of
the first kind. Therefore, only keeping the resonant terms, we can obtain the
final interaction Hamiltonian\CV
\begin{align}
H_{\text{I}}=&-\hbar\sum_{n=-N }^{N-1 }\sum_{m=-M}^{M}
 g_{x}a_{n+1,m}^{\dag}
a_{nm}+\text{H.c.} \nonumber\\ 
&+\hbar\sum_{n=-N }^{N }\sum_{m=-M}^{M-1}
  g_{y}e^{i\gamma n}a_{n,m+1}^{\dag}
a_{nm}+\text{H.c.}
,\label{eq:H_sp}%
\end{align}\CIV
where $g_{y}=$ $-2t_{y}\sin\left(  \frac{2\pi\bar{\Phi}}{\Phi_{0}}\right)
J_{1}\left(  \frac{2\pi\Phi_{\text{eff}}}{\Phi_{0}}\right)  $ is the \CV
column \CIV coupling strength. Here, we have concretely specified the length
(width) of the two-dimensional lattice $L=2N+1$ ($W=2M+1$). Equation
(\ref{eq:H_sp}) is similar to the Harper model%
~%
\cite{Hatsugai1993PRB,Hatsugai1993PRL} describing the two-dimensional integer
quantum Hall effect. 

\CV In Table.~\ref{tab:exp_pars}, we have listed typical values for the basic circuit parameters in experiment, where we have referred to Ref.~\cite{Geller2015PRA} for the capacitance and inductance values.
From these values, the other parameters, which we call the derived parameters, can be accordingly given in Table.~\ref{tab:quant_values}, either in the form of definite values or some numerical range. Beyond the range of the coupling strengths given in Table.~\ref{tab:quant_values}, we can choose more concrete values $g_x=g_y=2\pi\times4\operatorname{MHz}$, which can be actually achieved by applying proper bias magnetic flux for both row and column couplers.
\CIV
\begin{table}[ptbh]
	\caption{\CV  Typical values for basic circuit parameters. Below, $\Phi_0$ denotes the flux quantum. 
		\CIV}\label{tab:exp_pars}	
\begin{ruledtabular}
\begin{tabular}{lcc}
\textbf{ Basic circuit parameter}&\textbf{Symbol} & \textbf{Value} \\
\hline \\
 Equivalent junction inductance for odd rows & $L_{\mathrm{J},\mathrm{o}}$& $7.9\operatorname{nH}$  \\
 Equivalent junction inductance for even rows   &$L_{\mathrm{J},\mathrm{e}}$& $8.3\operatorname{nH}$  \\
 Coupler junction inductance &$L_{\mathrm{T}}$& $1.3\operatorname{nH}$\\
 Transmon shunting capacitance   &$C$& $ 91\operatorname{fF}$  \\
 Self inductance & $L_0$&  $210\operatorname{pH}$\\
 Mutual inductance & $M_0$ & $180\operatorname{pH}$\\
 
 Direct modulation amplitude of the bias &$\bar{\Phi}$&-$\frac{\Phi_0}{2}$ to $\frac{\Phi_0}{2}$\\
 \ \ magnetic flux for the column couplers &&\\
 Alternating modulation amplitude of the bias &$\Phi_\mathrm{eff}$ & -$\frac{\Phi_0}{2}$ to $\frac{\Phi_0}{2}$ \\
 \ \  magnetic flux for the column couplers&&\\
 Static bias magnetic flux for & $\Phi_\mathrm{o}$ & -$\frac{\Phi_0}{2}$ to $\frac{\Phi_0}{2}$\\
\ \  the odd-row couplers&&\\
 Static bias magnetic flux for& $\Phi_\mathrm{e}$ &-$\frac{\Phi_0}{2}$ to $\frac{\Phi_0}{2}$ \\
 \ \  the even-row couplers&&\\
Effective magnetic flux &$\gamma$& $-\pi$ to $\pi$
\end{tabular}
\end{ruledtabular}		
\end{table}

	\begin{table*}[tbp] \centering%
\caption{\CV Typical values for derived circuit parameters where the basic circuit parameters are specified according to Table.~\ref{tab:exp_pars}. In this table, $h$, $\hbar$, and $\Phi_0$ respectively mean the Plank constant, reduced Plank constant, and magnetic flux quantum. The function $J_1(x)$ is the Bessel function of the first kind.  Note that we should adjust $\Phi_{\text{o}}$ and $\Phi_{\text{e}}$ to make $G_\mathrm{o}=G_\mathrm{e}=-g_x$ and thus the row coupling strengths are homogeneous.\CIV}\label{tab:quant_values}%
\begin{ruledtabular}
\begin{tabular}
	[c]{llll}
	\textbf{Parameter} & \textbf{Symbol} & \textbf{Expression} & \textbf{Value}%
	\\\hline
	Odd-row Josephson energy & $E_{\text{J,o}}$ & $\frac{\Phi_{0}^{2}}{4\pi
		^{2}L_{\text{J,o}}}$ & $ 20.7 \operatorname{GHz} \cdot h$\\
	Even-row Josephson energy & $E_{\text{J,e}}$ & $\frac{\Phi_{0}^{2}}{4\pi
		^{2}L_{\text{J,e}}}$ & $ 19.7 \operatorname{GHz} \cdot h$ \\
	Charging energy & $E_{\text{C}}$ & $\frac{e^{2}}{2C}$ & $0.213 \operatorname{GHz} \cdot h$\\
	Odd-row plasma frequency & $\omega_{\text{p,o}}$ & $\frac{\sqrt{8E_{\text{C}%
			}E_{\text{J,o}}}}{\hbar}=\frac{1}{\sqrt{L_{\text{J,o}}C}}$ & $5.94 \operatorname{GHz}\cdot 2\pi$ \\
	Even-row plasma frequency & $\omega_{\text{p,e}}$ & $\frac{\sqrt{8E_{\text{C}%
			}E_{\text{J,e}}}}{\hbar}=\frac{1}{\sqrt{L_{\text{J,e}}C}}$ & $5.79 \operatorname{GHz}\cdot 2\pi$\\
	Odd-row qubit frequency & $\omega_{\text{o}}$ & $\omega_{\text{p,o}}%
	-\frac{E_{\text{C}}}{\hbar}$ & $5.72 \operatorname{GHz}\cdot 2\pi$\\
	Even-row qubit frequency & $\omega_{\text{e}}$ & $\omega_{\text{p,e}}%
	-\frac{E_{\text{C}}}{\hbar}$ & $5.58 \operatorname{GHz}\cdot 2\pi$\\
		Alternating modulation frequency for couplers along the columns & $\omega_{\mathrm{oe}}$ & $\omega_{\text{o}}-\omega_{\text{e}}$& $145 \operatorname{MHz}\cdot 2\pi$\\
	Odd-row qubit impedance & $Z_{\text{o}}$ & $\sqrt{\frac{L_{\text{J,o}}}{C}}$ & $ 295 \operatorname{\Omega}$
	\\
	Even-row qubit impedance & $Z_{\text{e}}$ & $\sqrt{\frac{L_{\text{J,e}}}{C}}$
	& $302 \operatorname{\Omega}$\\
	Bare column coupling strength & $t_{y}$ & $\frac{M_{0}^{2}\omega
		_{\text{p},\text{o}}\omega_{\text{p},\text{e}}}{4L_{\text{T}}\sqrt
		{Z_{\text{o}}Z_{\text{e}}}}$ & $4.51 \operatorname{MHz}\cdot 2\pi$\\
	Column coupling strength & $g_{y}$ & $-2t_{y}\sin\left(  \frac{2\pi\bar{\Phi}%
	}{\Phi_{0}}\right)  J_{1}\left(  \frac{2\pi\Phi_{\text{eff}}}{\Phi_{0}%
	}\right)  $ & $-5.25 \operatorname{MHz}\cdot 2\pi$ to $5.25 \operatorname{MHz}\cdot 2\pi$\\
	Mutual inductance induced by the odd-row couplers  & $M_{\text{o}}$ &
	$-\frac{M_{0}^{2}\cos\left(  \frac{2\pi}{\Phi_{0}}\Phi_{\text{o}}\right)
	}{L_{\text{T}}+2L_{0}\cos\left(  \frac{2\pi}{\Phi_{0}}\Phi_{\text{o}}\right)
	}$ & $-18.8 \operatorname{pH}$ to $36.8 \operatorname{pH}$\\
	Mutual inductance induced by the even-row couplers & $M_{\text{e}}$ & 
	$-\frac{M_{0}^{2}\cos\left(  \frac{2\pi}{\Phi_{0}}\Phi_{\text{e}}\right)
	}{L_{\text{T}}+2L_{0}\cos\left(  \frac{2\pi}{\Phi_{0}}\Phi_{\text{e}}\right)
	}$ & $-18.8 \operatorname{pH}$ to $36.8 \operatorname{pH}$\\
	Odd-row couplings strength & $G_{\text{o}}$ & $-\frac{M_{\text{o}}}%
	{2}\frac{\omega_{\text{p,o}}^{2}}{Z_{\text{o}}}$ & $-13.8 \operatorname{MHz}\cdot 2\pi$ to $7.08 \operatorname{MHz}\cdot 2\pi$\\
	Even-row couplings strength & $G_{\text{e}}$ & $-\frac{M_{\text{e}}}%
	{2}\frac{\omega_{\text{p,e}}^{2}}{Z_{\text{e}}}$ & $-12.8 \operatorname{MHz}\cdot 2\pi$ to $6.57 \operatorname{MHz}\cdot 2\pi$\\
\end{tabular}%
\end{ruledtabular}
\end{table*}%

\subsection{Quasimomentum-space Hamiltonian}%

\CV
Just as in the two-dimensional topological physics%
~%
\cite{Hatsugai1993PRL,Hatsugai1993PRB}, we now prefer to investigate one
dimension in the quasimomentum representation and the other one in the lattice
representation. Here, the quasimomentum means the wave vector regarding the
single-particle Bloch function. Accordingly, the quasimomentum (lattice)
representation means the second quantization using the Bloch (Wannier)
function basis. We mention that the Hamiltonian constructed by a qubit system
is intrinsically of the lattice representation, or tight-binding form [see Eq.%
~%
(\ref{eq:H_sp})].

Here, we will enter the quasimomentum representation for the row direction,
but remain in the lattice representation for the column direction. To do this,
we make the infinite length ($L=\infty$) assumption for Eq.%
~%
(\ref{eq:H_sp}), and then make the transformation \ \
\begin{equation}
a_{nm}=\frac{1}{\sqrt{L}}e^{i\gamma mn}\sum_{k_{x}}e^{ik_{x}n}b_{k_{x}%
,m},\label{eq:sigma-S}%
\end{equation}
where the presence of $e^{i\gamma mn}$ is used to compensate the effect of
$g_{y}e^{i\gamma n}$ in Eq.%
~%
(\ref{eq:H_sp}) which breaks the lattice translation invariance of the
Hamiltonian along the row direction ($n\rightarrow n+1$), and besides,
$b_{k_{x},m}$ is the collective annihilation operator at the row-direction
quasimomentum $k_{x}$ and column-direction location $m$. After the
transformation in Eq.%
~%
(\ref{eq:sigma-S}), the Hamiltonian $H_{\text{I}}$ in Eq.%
~%
(\ref{eq:H_sp}) is changed into what we call the quasimomentum-representation
Hamiltonian
\begin{equation}
H_{\text{I}}=\sum_{k_{x}}\tilde{H}\left(  k_{x}\right)  ,\label{eq:HI_QMR}%
\end{equation}
where
\begin{align}
\tilde{H}\left(  k_{x}\right)   &  =\sum_{m=-M}^{M}\hbar\varepsilon_{\gamma
m+k_{x}}b_{k_{x},m}^{\dag}b_{k_{x},m}\nonumber\\
&  +\sum_{m=-M}^{M-1}\left(  \hbar g_{y}b_{k_{x},m+1}^{\dag}b_{k_{x}%
,m}+\text{H.c.}\right)  .\label{eq:H_k}%
\end{align}
with the notation $\varepsilon_{k}=-2g_{x}\cos k~$is defined as the
quasimomentum-space Hamiltonian. Here, the coupling strengths are generally of
the order $g_{x\left(  y\right)  }\sim$ $2\pi\times4%
\operatorname{MHz}%
$%
~%
\cite{Roushan2017NP}. It can be obviously seen from Eq.%
~%
(\ref{eq:HI_QMR}) that the interactions along the row direction are decoupled
after the transformation in Eq.%
~%
(\ref{eq:sigma-S}).%

\CIV
%

\CV

\section{Three-leg model}%

\CIV

\label{sec:phase transition}

\subsection{Single-particle energy spectrum}

\label{sec:single-particle energy spectrum}

\begin{figure}[ptb]
\centering\includegraphics[width=0.48\textwidth,clip]{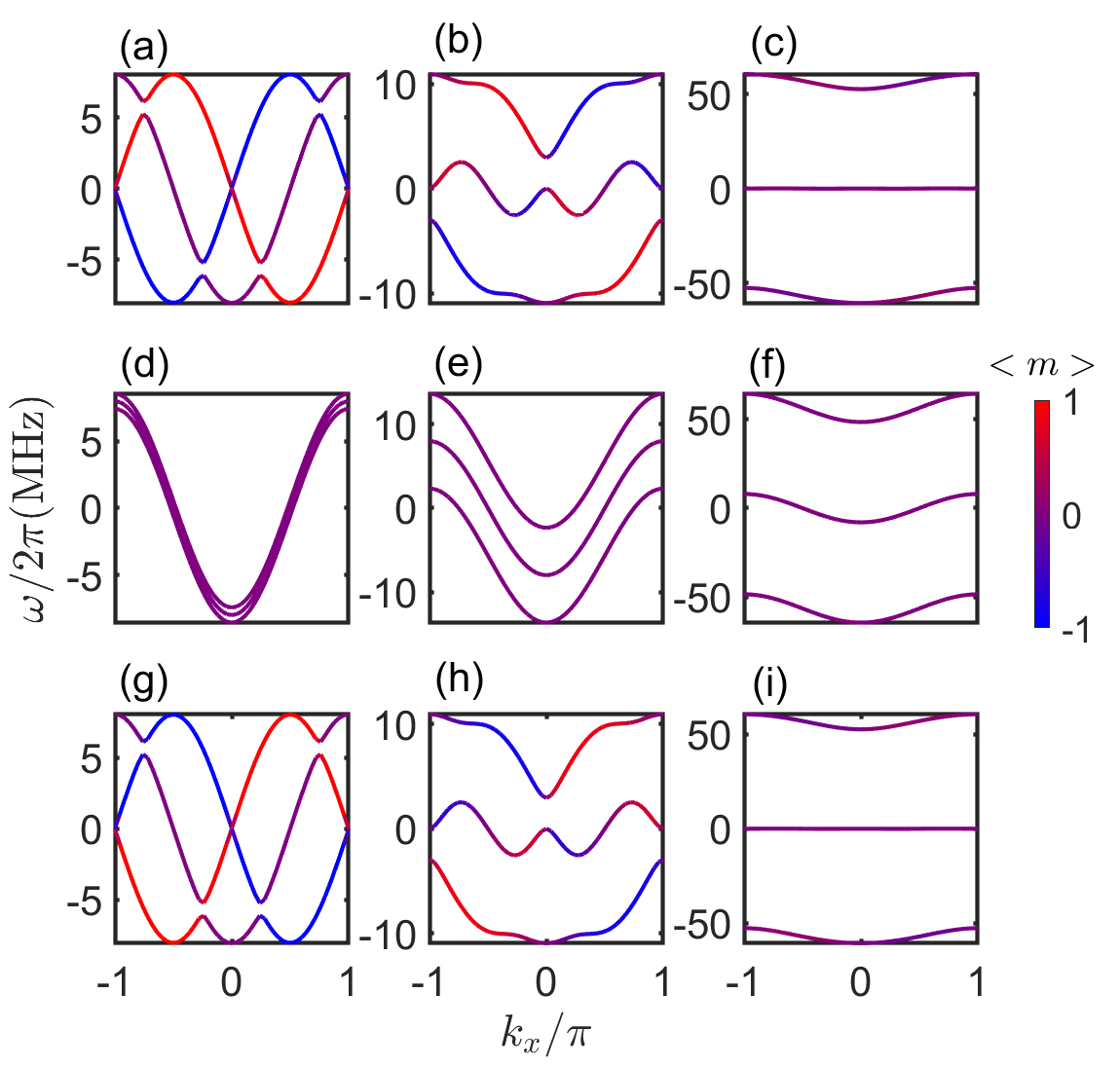}\caption{(color
online) Energy bands for the double-ladder model \color{black}(i.e., the
lattice width $W=3$), where $\omega$ ($k_{x}$) denotes the eigenfrequency
(row-direction quasimomentum)\color{black}. Besides, the row coupling strength
takes $g_{x}=2\pi\times4\operatorname{MHz}$ in all plots. However, in each
column from leftmost to rightmost, the column coupling strength takes
$g_{y}=0.1g_{x}$, $g_{x}$, and $10g_{x}$, respectively. And in each row from
top to bottom, the effective magnetic flux takes $\gamma=-\frac{\pi}{2}$, $0$,
and $\frac{\pi}{2}$, respectively. Furthermore, the curve color represents
$\left\langle m\right\rangle $, i.e., the average location along the $y$
direction in the present state. }%
\label{fig:ChiralCurrent}%
\end{figure}

Firstly, we focus on the special double-ladder model, i.e., the simplest
two-dimensional Harper model with the lattice width taking $W=3$. Then, from
Eq.%
~%
(\ref{eq:H_k}), the quasimomentum-space Hamiltonian can be represented in the
concise form $\tilde{H}\left(  k_{x}\right)  =b_{k_{x}}^{\dag}h\left(
k_{x}\right)  b_{k_{x}}$ where $b_{k_{x}}=\left(  b_{k_{x},-1},b_{k_{x}%
,0},b_{k_{x},1}\right)  ^{\intercal}$ is an operator vector and the
single-particle Hamiltonian $h\left(  k_{x}\right)  =h_{z}\left(
k_{x}\right)  +h_{x}$ is a $3\times3$ matrix with%
\begin{align}
h_{z}\left(  k_{x}\right)   &  =\hbar%
\begin{pmatrix}
\varepsilon_{k_{x}-\gamma} & 0 & 0\\
0 & \varepsilon_{k_{x}} & 0\\
0 & 0 & \varepsilon_{k_{x}+\gamma}%
\end{pmatrix}
,\\
h_{x}  &  =\hbar g_{y}%
\begin{pmatrix}
0 & 1 & 0\\
1 & 0 & 1\\
0 & 1 & 0
\end{pmatrix}
.\label{eq:hzhx}%
\end{align}
Using the superscript $\intercal$ to denote the matrix transposition, the
eigenvectors of $h_{z}$ are then $e_{z,-1}=\left(  1,0,0\right)  ^{\intercal}%
$, $e_{z,0}=\left(  0,1,0\right)  ^{\intercal}$, and $e_{z,1}=\left(
0,0,1\right)  ^{\intercal}$ with eigenvalues $E_{z,-1}=\hbar\varepsilon
_{k_{x}-\gamma}$, $E_{z,0}=\hbar\varepsilon_{k_{x}}$, and $E_{z,1}%
=\hbar\varepsilon_{k_{x}+\gamma}$, respectively. However, an exception occurs
for $\gamma=0$, when $h_{z}=\hbar\varepsilon_{k_{x}}I$ with $I$ being the
identity matrix. The eigenvectors of $h_{x}$ are $e_{x,-1}=\frac{1}{2}\left(
1,-\sqrt{2},1\right)  ^{\intercal}$, $e_{x,0}=\frac{1}{2}\left(  -\sqrt
{2},0,\sqrt{2}\right)  ^{\intercal}$, and $e_{x,1}=\frac{1}{2}\left(
1,\sqrt{2},1\right)  ^{\intercal}$ with eigenvalues $E_{x,-1}=-\sqrt{2}\hbar
g_{y}$, $E_{x,0}=0$, and $E_{x,1}=\sqrt{2}\hbar g_{y}$. The competition
between $h_{z}$ and $h_{x}$ results in the energy bands\ shown in Fig.%
~%
\ref{fig:ChiralCurrent}, where $g_{x}$ is fixed at $2\pi\times4%
\operatorname{MHz}%
$, but $g_{y}$ and $\gamma$ are varied.

For $\gamma=-\frac{\pi}{2}$ and $g_{y}=0.1g_{x}$ [see Fig.%
~%
\ref{fig:ChiralCurrent}(a)], the energy bands are mainly determined by $h_{z}
$, which possesses three-branch eigenvalues $\hbar\varepsilon_{k_{x}-\gamma}$,
$\hbar\varepsilon_{k_{x}}$, and $\hbar\varepsilon_{k_{x}+\gamma}$
corresponding to eigenvectors $e_{z,-1}$, $e_{z,0}$, and $e_{z,1}$. Under each
vector, the average location along the column direction is accordingly
$\left\langle m\right\rangle _{e_{z,-1}}=-1$, $\left\langle m\right\rangle
_{e_{z,0}}=0$ and $\left\langle m\right\rangle _{e_{z,1}}=1$. In addition, the
minimum points of these three branches are degenerate at $k_{x}=\gamma,0$, and
$-\gamma$, respectively. However, the presence of $h_{x}$ makes these three
branches hybridized, ending with a broken degeneracy. Then, the minimum value
is actually only achieved at $k_{x}=0$ [see Appendix.%
~%
\ref{Append:Degen}]. When $g_{y}$ is increased to $g_{x}$ [see Fig.%
~%
\ref{fig:ChiralCurrent}(b)], $h_{x}$ becomes more important, the unique
minimum at $k_{x}=0$ can be figured out more apparently. When $g_{y}=10g_{x}$
[see Fig.%
~%
\ref{fig:ChiralCurrent}(c)], $h_{x}$ nearly dominates the single-particle
Hamiltonian. Under the eigenvectors of $h_{x}$, the average location along the
column direction takes $\left\langle m\right\rangle _{e_{x,j}}=0~$($j=1,2,3$).
The three energy bands become flatter, which are $E_{x,-1}^{\prime}=-\sqrt
{2}\hbar g_{y}+\frac{\hbar}{4}\varepsilon_{k_{x}-\gamma}+\frac{\hbar}%
{2}\varepsilon_{k_{x}}+\frac{\hbar}{4}\varepsilon_{k_{x}+\gamma}$,
$E_{x,0}^{\prime}=\frac{\hbar}{2}\varepsilon_{k_{x}-\gamma}+\frac{\hbar}%
{2}\varepsilon_{k_{x}+\gamma}$, and $E_{x,1}^{\prime}=\sqrt{2}\hbar
g_{y}+\frac{\hbar}{4}\varepsilon_{k_{x}-\gamma}+\frac{\hbar}{2}\varepsilon
_{k_{x}}+\frac{\hbar}{4}\varepsilon_{k_{x}+\gamma}$ by the perturbative
theory. Besides, there is still a unique minimum point which only occurs at
$k_{x}=0$.

When we change $\gamma$ to $0$, $h_{z}\,=\hbar\varepsilon_{k_{x}}$ is in fact
a global shift to $h_{x}$. Since the eigenenergies of $h_{x}$ are independent
of $k_{x}$, the energy gaps maintain constant for changing $k_{x} $ [see Figs.%
~%
\ref{fig:ChiralCurrent}(d)-\ref{fig:ChiralCurrent}(f)]. But there is a band
bending when $k_{x}$ varies, which\ is fundamentally induced by the function
$\varepsilon_{k_{x}}$. When $g_{y}=10g_{x}$, the bending becomes less obvious
[see Fig.%
~%
\ref{fig:ChiralCurrent}(f)] because $h_{x}$ becomes dominant. The three bands
still correspond to the eigenstates $e_{xj}$ ($j=1,2,3$), under which, the
average location along the column direction takes $\left\langle m\right\rangle
_{e_{x,j}}=0~$($j=1,2,3$).

Next, we focus on the case $\gamma=\frac{\pi}{2}$, where the single-particle
Hamiltonian $\left.  h_{k_{x}}\right\vert _{\gamma=\frac{\pi}{2}}$ can be
easily proved to be equivalent to $\left.  h_{-k_{x}}\right\vert
_{\gamma=-\frac{\pi}{2}}$ [see Eq.%
~%
(\ref{eq:hzhx})]. Thus, the energy bands in Figs.%
~%
\ref{fig:ChiralCurrent}(g)-\ref{fig:ChiralCurrent}(i) can be obtained by
symmetrizing Figs.%
~%
\ref{fig:ChiralCurrent}(a)-\ref{fig:ChiralCurrent}(c) with respect to the axis
$k_{x}=0$. Hence, we \CV just skip making \CIV further discussions for this case.

\subsection{Current pattern in the open-boundary condition}

\label{Sec:CurrentPattern}%

\begin{figure*}[ptbh]
\centering\includegraphics[
width=0.9\textwidth,clip
]{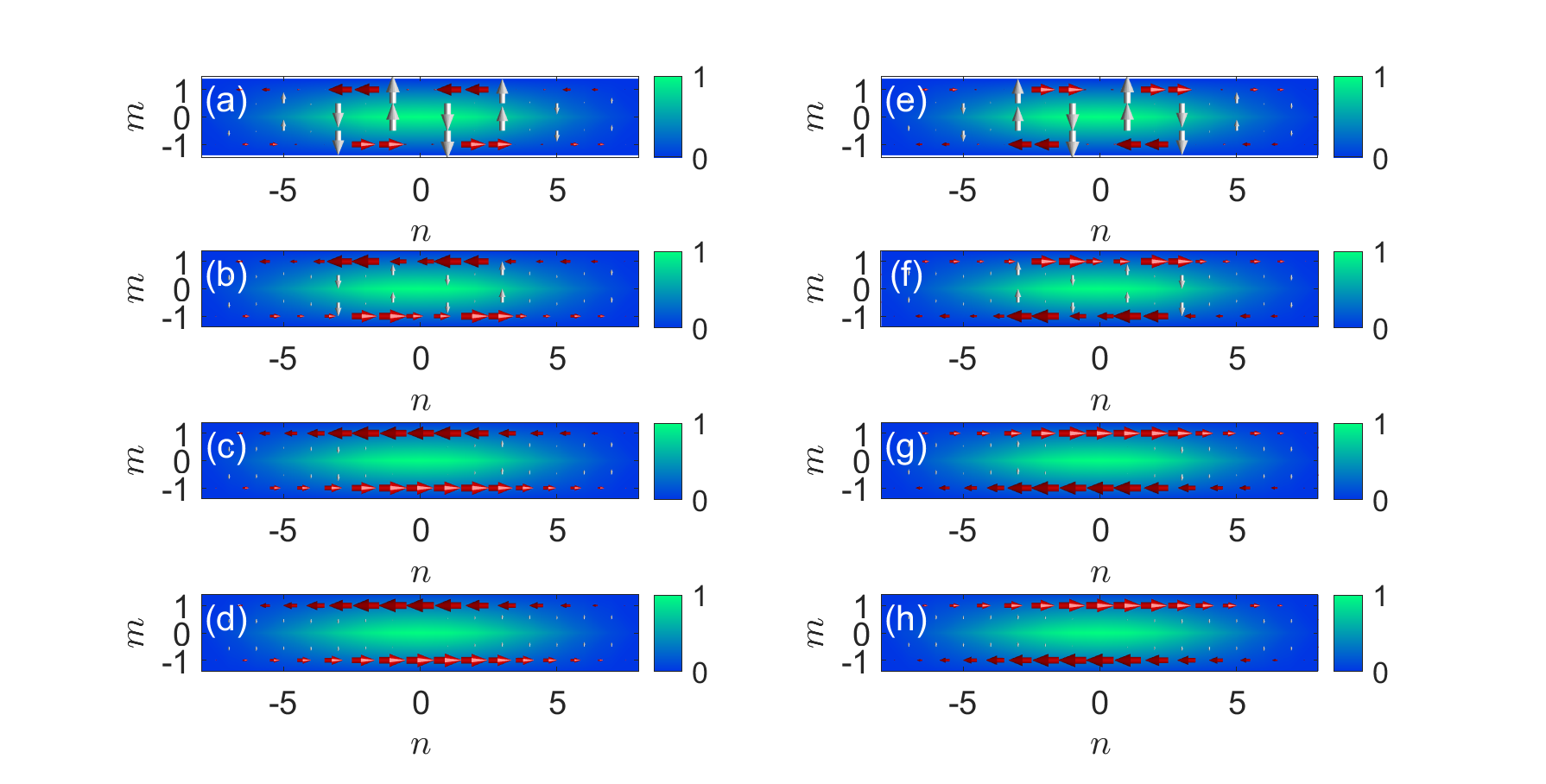}
\caption
{(color online) Particle current patterns for the effective magnetic flux $\gamma
$ taking (a)-(c) $-\frac{\pi}{2}$
and (d)-(f) $\frac{\pi}{2}%
$ in a \CV three-leg \CIV ladder (\CV lattice \CIV width $W=3$) with  the \CV lattice \CIV length $L=17$. From top to bottom, the coupling strengths along two directions fulfill
the conditions $K=0.1$, $0.2$, $0.4$, and $0.7$, respectively. The particle current between adjacent sites is represented
by a arrow whose size implies the current magnitude.
We have used the red (green) color for currents along the edge (in the bulk).
Beside, the color in the background represents the relative occupation probability in the
single-particle ground state. (See Fig.~\ref{fig:Vortex-Meissner-Normalized}
in Appendix.~\ref{Append:NormalizedCurrentPattern}
for normalized current patterns where the number of vortices can be counted more easily.)}
\label{fig:Vortex-Meissner}
\end{figure*}%

In the open-boundary condition, the continuous spectrum in Fig.%
~%
\ref{fig:ChiralCurrent} is discretized. We assume that the single-particle
eigenstate regarding the energy level $E_{j}$ ($j=1,2,\cdots,WL$, and
$E_{1}\leq E_{2}\cdots\leq E_{WL})$ is $\left\vert E_{j}\right\rangle
=S_{j}^{+}\left\vert 0\right\rangle $, where $\left\vert 0\right\rangle $ is
the (global) ground state and
\begin{equation}
S_{j}^{+}=\sum_{n=-N}^{N}\sum_{m=-M}^{M}\psi_{nm}^{\left(  j\right)  }%
a_{nm}^{\dag}\label{eq:Sigma_j_p}%
\end{equation}
is single-particle eigenstate creation operator. Particularly, the
single-particle ground state is namely $\left\vert G\right\rangle =\left\vert
E_{1}\right\rangle =S_{1}^{+}\left\vert 0\right\rangle $.

To characterize the current pattern in the single-particle ground state, we
first seek $I_{nm;n^{\prime}m^{\prime}}$, the particle current operator from
the site $nm$ to $n^{\prime}m^{\prime}$, which, from the continuity equation%
\begin{equation}
\frac{\text{d}}{\text{d}t}\left(  a_{nm}^{\dag}a_{nm}\right)  =\frac{\left[
a_{nm}^{\dag}a_{nm},H\right]  }{i\hbar}=-\sum_{\,\text{adjacent }n^{\prime
}m^{\prime}}I_{nm;n^{\prime}m^{\prime}}%
\end{equation}
can be derived to possess the general form as follows, that is,
\begin{equation}
I_{nm;n^{\prime}m^{\prime}}=-i\left(  g_{nm;n^{\prime}m^{\prime}}%
a_{nm}a_{n^{\prime}m^{\prime}}^{\dag}-\mathrm{H.c.}\right)  \text{.}%
\label{eq:Inmnpmp}%
\end{equation}
Here, $g_{nm;n^{\prime}m^{\prime}}$ represents the coupling strength between
the sites $nm$ and $n^{\prime}m^{\prime}$ in the original Hamiltonian $H$ [see
Eq.%
~%
(\ref{eq:H_sp})]. For example, for $\left(  n^{\prime},m^{\prime}\right)  $
taking $\left(  n+1,m\right)  $, $g_{nm;n^{\prime}m^{\prime}}$ is replaced
with $-g_{x}$, while for $\left(  n^{\prime},m^{\prime}\right)  $ taking
$\left(  n,m+1\right)  $, $g_{nm;n^{\prime}m^{\prime}}$ is replaced with
$g_{y}e^{i\gamma n}$. In the state $\left\vert G\right\rangle $, the mean
current from the site $nm$ to $n^{\prime}m^{\prime}$ is then
\begin{align}
I_{nm;n^{\prime}m^{\prime}}^{\text{G}}  &  =\left\langle G\right\vert
I_{nm;n^{\prime}m^{\prime}}\left\vert G\right\rangle \nonumber\\
&  =-i\left(  g_{nm;n^{\prime}m^{\prime}}\psi_{nm}^{\left(  1\right)  }%
\psi_{n^{\prime}m^{\prime}}^{\left(  1\right)  \ast}-\mathrm{c.c.}\right)
.\label{eq:ChiralCurrent}%
\end{align}

Before intuitively showing the current patterns, we first introduce the
concept of vortex, which represents a current pattern that all the particle
currents circulate around a center either clockwise or anticlockwise. In Fig.%
~%
\ref{fig:Vortex-Meissner}, the current patterns have been plotted for
different effective magnetic fluxes $\gamma$ and coupling ratios
$K=g_{y}/g_{x}$. For the case of $\gamma=$ $-{\pi}/{2}$ [see Figs.~\ref{fig:Vortex-Meissner}(a)-(d)], we can see the vortex number continues to
decrease, in detail, from 7, 4, 2, to 1, for $K$ taking 0.1, 0.2, 0.4, and
0.7, respectively. 
\CV In analogue to the vortex-Meissner phase transition in superconductor
material \CV and also in the two-leg ladder model~\cite{Zhao2020PRA,Atala2014NP}, here we will continue to call a single vortex (e.g, Figs.~\ref{fig:Vortex-Meissner}(d)) as ``Meissner phase'' but multiple vortices (Figs.~\ref{fig:Vortex-Meissner}(a)-(c)) as ``vortex phase'' although it will be shown in the following that no phase transition by definition has actually happened. \CIV 
If the $\gamma$ is flipped to ${\pi}/{2}$ [see Figs.%
~%
\ref{fig:Vortex-Meissner}(\CV e\CIV)-\ref{fig:Vortex-Meissner}(\CV h\CIV)], we see the
particle currents only changes the directions for the same coupling strength
ratio. This reveals the coupling ratio is critical to the \CV transition between ``vortex phase''
 and ``Meissner phase''\CIV. As illustrated by the background color in each panel, the
bulk (i.e., the central row) is mostly populated compared to edges, which,
however, does not induce a larger particle current from or to the central row
sites, in that the particle current is additionally affected by the relative
phase between adjacent sites.

For example, in Fig.%
~%
\ref{fig:Vortex-Meissner}, the particle current in the bulk, i.e., the central
row, is always zero. This can be explained by investigating the property of
the single-particle ground state $\psi_{nm}^{\left(  1\right)  }\equiv
\psi_{nm}^{\left(  1\right)  }\left(  \gamma\right)  $. To do this, we first
consider the case of broken time-reversal symmetry ($\gamma\neq0,\pm\pi$).
Then we rotate the double ladder by $\pi$ with respect the axis $m=0$, after
which, the Hamiltonian $H_{\text{I}}\equiv H_{\text{I}}\left(  \gamma\right)
$ and wave function $\psi_{nm}^{\left(  1\right)  }\left(  \gamma\right)  $
respectively become $H_{\text{I}}\left(  -\gamma\right)  $ and $\psi
_{n,-m}^{\left(  1\right)  }\left(  \gamma\right)  $, where the relation
$H_{\text{I}}\left(  -\gamma\right)  =E_{1}\psi_{n,-m}^{\left(  1\right)
}\left(  \gamma\right)  $ must hold. On the other hand, we note $H_{\text{I}%
}\left(  -\gamma\right)  $ can also be obtained by taking the time-reversal of
$H_{\text{I}}\left(  \gamma\right)  $ and the eigenstate of $H_{\text{I}%
}\left(  -\gamma\right)  $ with the eigen energy $E_{1}$ should be equivalent
to $\psi_{n,m}^{\left(  1\right)  \ast}\left(  \gamma\right)  $ (see Appendix.%
~%
\ref{Append:Degen} ). Thus, via a well-chosen gauge, we can simply think
$\psi_{n,m}^{\left(  1\right)  \ast}\left(  \gamma\right)  \equiv\psi
_{n,-m}^{\left(  1\right)  }\left(  \gamma\right)  $, which further gives
$\psi_{n,0}^{\left(  1\right)  \ast}\left(  \gamma\right)  \equiv$ $\psi
_{n,0}^{\left(  1\right)  }\left(  \gamma\right)  $. This means the wave
function component $\psi_{n,0}^{\left(  1\right)  }\left(  \gamma\right)  $
must be real. From the particle-current formula [see Eq.%
~%
(\ref{eq:ChiralCurrent})], we obtain the immediate result $I_{n,0;n+1,0}%
^{\text{\textrm{G}}}=0$, the very revelation of the zero particle current in
the row $m=0$. If the time-reversal symmetry is conserved ($\gamma=0,\pm\pi$),
the matrix form of $H_{\text{I}}$ is real and symmetrical, and then all the
eigenstates of $H_{\text{I}}$ must be real vectors, e.g., $\psi_{n,0}^{\left(
1\right)  \ast}\left(  \gamma\right)  \equiv$ $\psi_{n,0}^{\left(  1\right)
}\left(  \gamma\right)  $. This also implies zero particle current along the
central row. In contrast to the two-leg ladder%
~%
\cite{Zhao2020PRA,Atala2014NP}, the double (or three-leg) ladder is the
minimal configuration for which chiral currents at the edges can be sharply
distinguished from the behavior of the bulk%
~%
\cite{Mancini2015Science}.

\subsection{Vortex number}%

\begin{figure}[ptbh]
\centering\includegraphics[width=0.5\textwidth,clip]{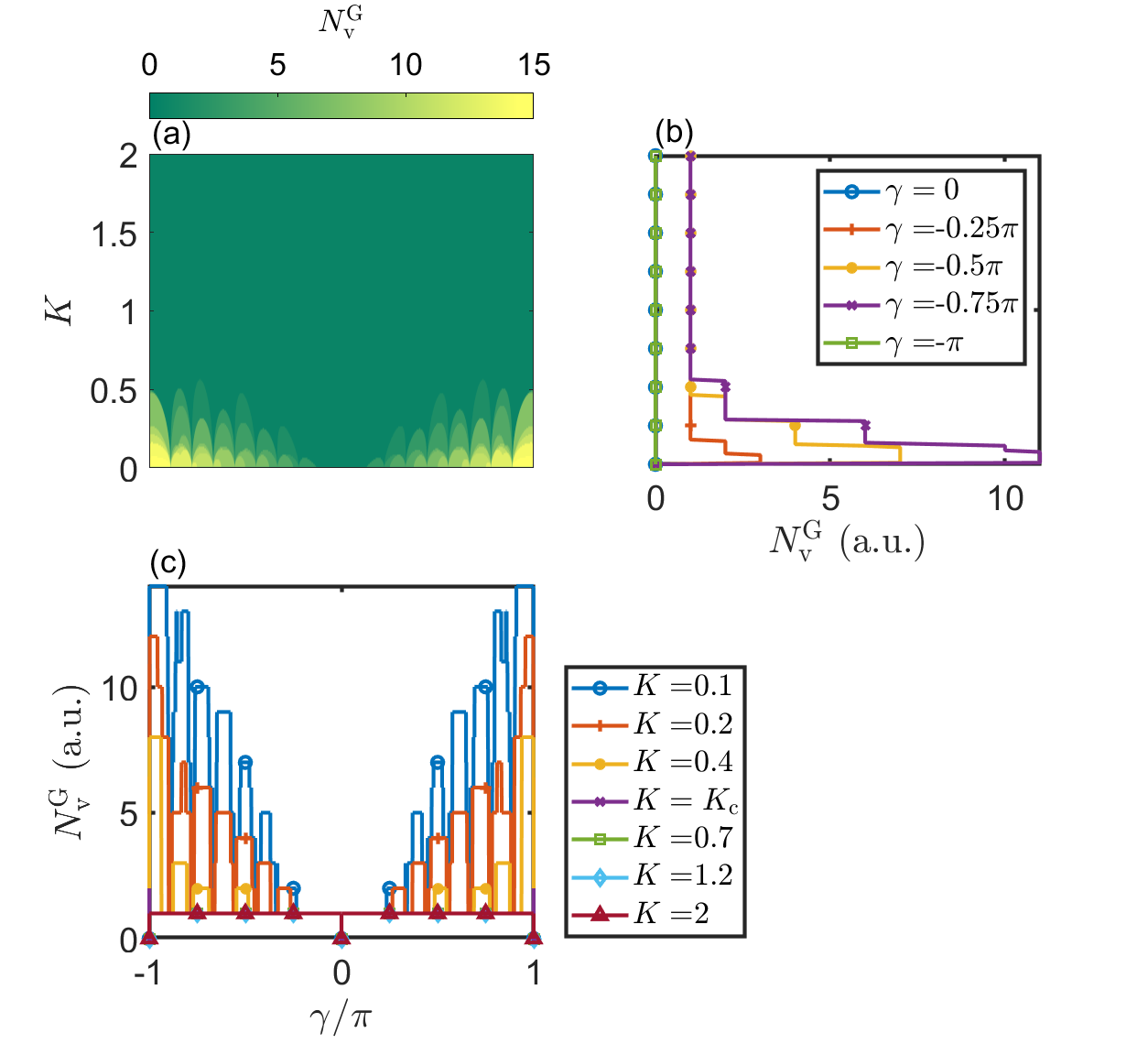}\caption{(color
online) Vortex number \CV$N_{\text{v}}^{\text{G}}$ \CIV
plotted (a) against the effective magnetic
flux $\gamma$ and coupling ratio $K$, (b) against $K$ for different $\gamma
$, and
(c) against $\gamma$ for different $K$. In (b), the curve for $\gamma
=0$ is covered
by that for $\gamma=\pi
$, while in (c), the curves for $K=0.7$ and $1.2$ are covered
by that for $K=2$. Here, $K_c=0.55$. The lattice width (length) is $W=3$ ($L=17$).}
\label{fig:PlotVortexDensity}
\end{figure}%

Here, we will introduce the finite-size effect that is not reported in the
two-leg ladder%
~%
\cite{Atala2014NP,Zhao2020PRA}\CV, which will further induce what we call the ``staggered vortex-Meissner phase transition. \CIV In Sec.%
~%
\ref{sec:single-particle energy spectrum}, we have demonstrated there is
always one minimum point on the energy bands, corresponding to the unique
single-particle ground state $b_{k_{x}=0}^{\dag}\left\vert 0\right\rangle $
[see Eq.%
~%
(\ref{eq:sigma-S})], where $\left\vert 0\right\rangle $ is the global ground
state. If the lattice length $L$ approaches infinity, we can approximately
think the single-particle ground state in the open-boundary condition is
$\left\vert G\right\rangle =b_{k_{x}=0}^{\dag}\left\vert 0\right\rangle $,
whose wave function is therefore $\psi_{nm}^{\left(  1\right)  }=e^{i\gamma
mn}e_{m}\left(  0\right)  /\sqrt{L} $. Here, $\left(  e_{-1}\left(  0\right)
,e_{0}\left(  0\right)  ,e_{1}\left(  0\right)  \right)  ^{\intercal}$ is the
eigen vector of $\left.  h\left(  k_{x}\right)  \right\vert _{k_{x}=0}$ with
the smallest eigen value. Then, the particle current along the column
direction can be calculated as
\begin{align}
I_{n,-1;n,0}^{\text{G}}  &  =I_{n,0;n,1}^{\text{G}}\equiv I_{n}^{\text{G}%
}\nonumber\\
&  =-ig_{y}\psi_{n,0}^{\left(  1\right)  }\left(  e^{i\gamma n}\psi
_{n,-1}^{\left(  1\right)  }-\mathrm{c.c.}\right) \nonumber\\
&  =0.\label{eq:In}%
\end{align}
Simultaneously, the particle current along the edge rows are $I_{n,-1,;n+1,-1}%
^{\text{\textrm{G}}}=-I_{n,1;n+1,1}^{\text{\textrm{G}}}=-({2g_{x}}/
{L})\left\vert e_{-1}\left(  0\right)  \right\vert ^{2}\sin\gamma$. This means
there is always one mere big vortex. However, the practical length $L$ can not
be infinite. In the below, we will show for a practical finite length $L$, the
vortex number can exhibit rich phenomena according to different effective
magnetic flux $\gamma$ and coupling ratio $K$.

Now, we investigate how the vortex number changes against the effective
magnetic flux $\gamma$ and coupling ratio $K$. As shown in Fig.%
~%
\ref{fig:PlotVortexDensity}(a), the vortex number is symmetric about the axis
$\gamma=0$. In addition, the vortex number exceeding one merely occurs on left
and right bottom corners, where one also find some discrepancies with zero
vortex numbers. In Fig.%
~%
\ref{fig:PlotVortexDensity}(b), the vortex number is plotted against $K$ for
$\gamma=0$, $-0.25\pi$, $-0.5\pi$, $-0.75\pi$, and $-\pi$, respectively. It is
clearly shown that the vortex number decreases as $K$ becomes larger. Also,
there exists a threshold of $K$ for each specified $\gamma$, above which, the
vortex number will remain one steadily. In Fig.%
~%
\ref{fig:PlotVortexDensity}(c), the vortex number is plotted against $\gamma$
for $K=0.1$, $0.2$, $0.4$, $K_{\text{c}}$, 0.7, 1.2, and $2$, respectively.
Here, $K_{\text{c}}$ is the global threshold of $K$, above which the vortex
number becomes one whatever $\gamma$ takes. In the present case $N=17$,
$K_{\text{c}}=0.55$ can be numerically obtained. As in Fig.%
~%
\ref{fig:PlotVortexDensity}(a), we also see in Fig.%
~%
\ref{fig:PlotVortexDensity}(c) the symmetry about $\gamma=0$. When $\left\vert
\gamma\right\vert $ approaches zero, we see the vortex number first drops in
 \CV a staggered \CIV manner and then remains one for $\gamma$ exceeding a particular
threshold. Before this threshold is met, there are pieces of intervals along
the $\gamma$ axis where only one vortex occurs, which however are never
reported in the case of two-leg ladder%
~%
\cite{Atala2014NP,Zhao2020PRA}. \CV Here, we call this kind of behavior of the vortex number as ``staggered vortex-Meissner phase transition'' considering that a single vortex and multiple vortices have already been defined as ``Meissner phase'' and ``vortex phase'' respectively in analogue to the vortex-Meissner phase transition in superconductor
material and also in the two-leg ladder~\cite{Zhao2020PRA,Atala2014NP}. We can also find that there \CIV exists the global threshold of
$\gamma$, above which the vortex number always remains one whatever $K$ takes.
In this case of $L=17$, we can obtain this global threshold is $\gamma
_{\text{c}}=0.14\pi$. As previously mentioned, at the time-reversal symmetric
points ($\gamma=0,\pm\pi$), the energy eigenstates are purely real vectors,
and thus there is no particle currents according to Eq.%
~%
(\ref{eq:ChiralCurrent}), which is why we observe zero vortex number at these
points in Fig.%
~%
\ref{fig:PlotVortexDensity}(c).

\begin{figure}[tb]
	\centering
	\includegraphics[width=0.23\textwidth]{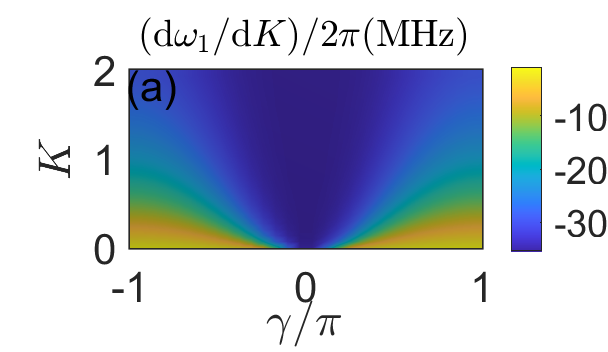}
	\includegraphics[width=0.23\textwidth]{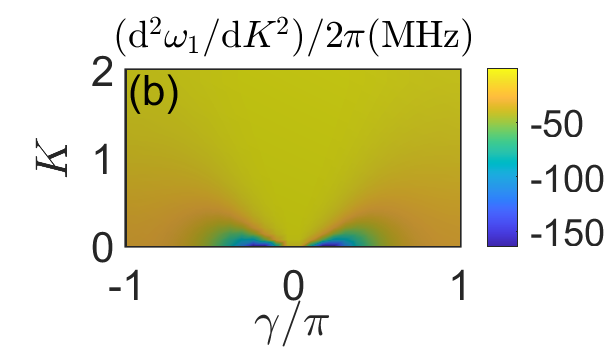}\\
	\includegraphics[width=0.23\textwidth]{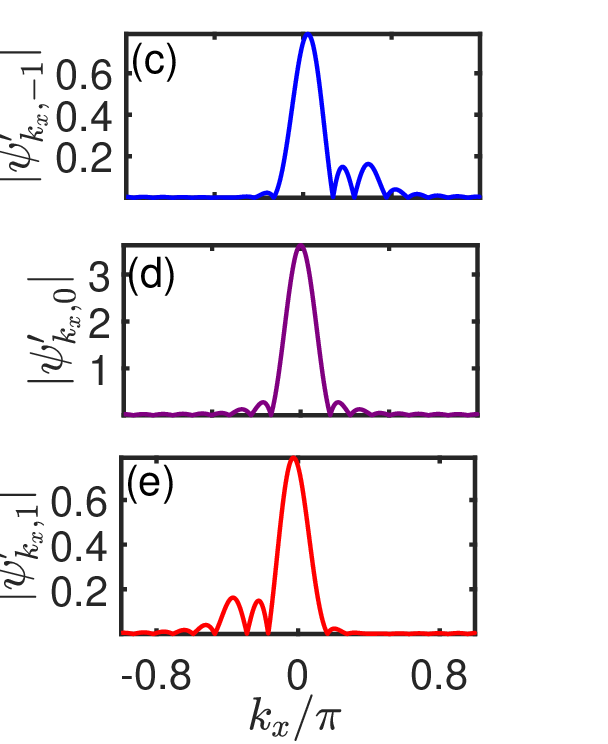}	\includegraphics[width=0.23\textwidth]{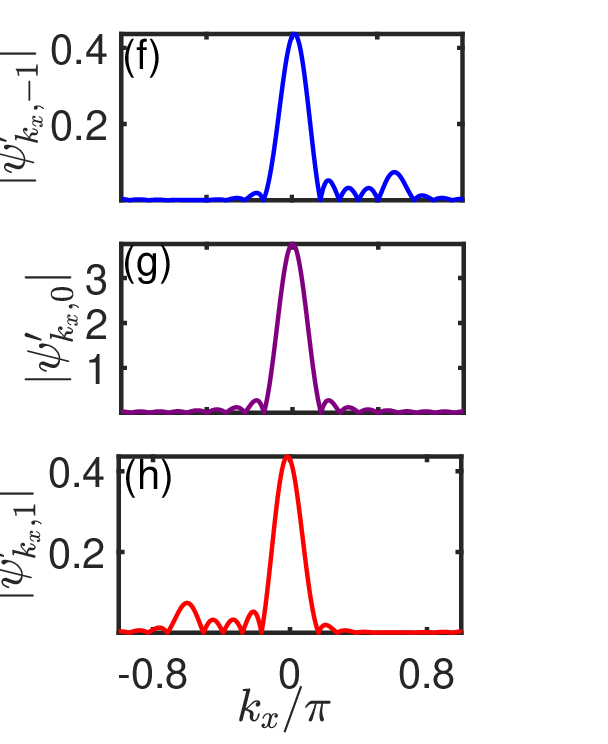}
	\caption{\CV (a ) First-order and (b) second-order derivatives of $\omega_1$ with respect to the coupling ratio $K$, against $K$ and $\gamma$. In (c)-(e) [(f)-(h)], we have shown quasimomentum distribution the single-particle-ground-state wave function for the row $m=-1,0,1$, denoted by $|\psi'_{k_x,-1}|$, $|\psi'_{k_x,0}|$, and $|\psi'_{k_x,1}|$ respectively for the coupling ratio $K=0.3$ and effective magnetic flux $\gamma=0.4$ ($\gamma=0.6$) where there should be one vortex (three vortices) according to Fig.~\ref{fig:PlotVortexDensity}(a). In addition, the row coupling strength $g_x/2\pi=4\operatorname{MHZ}$, the lattice with $W=3$, and lattice length $L=17$. \CIV \color{red}\color{black}\label{fig:phase_transition}}
\end{figure}

\CV
However, we must point out that the ``staggered vortex-Meissner phase transition'' here instead of real phase transition by definition, is just a similar description for the vortex number transition between one and larger integers as in the vortex-Meissner phase transition in superconductor
material and also in the two-leg ladder~\cite{Zhao2020PRA,Atala2014NP}. Based on Landau's phase-transition theory and the three-leg ladder single-particle bulk-sate energy spectrum [e.g., Figs.~\ref{fig:ChiralCurrent}(a)-2(c)], we are motivated to regard the single-particle-state energy $\hbar \omega$ as the ``free energy'', the coupling ratio $K$ as the ``temperature'', and the quasimomentum $k_x$ as the ``order parameter''. Thus, the phase transition is ready to be discussed. We have numerically given both the first-order [see Figs.~\ref{fig:phase_transition}(a)] and second-order [see Figs.~\ref{fig:phase_transition}(b)] derivatives of the single-particle ground-state eigenfrequency $\omega_1$ in the open-boundary conditions with respect to the coupling ratio $K$ against $K$ and the effective magnetic flux $\gamma$, where the row coupling strength $g_x/2\pi=4\operatorname{MHZ}$, the lattice with $W=3$, and lattice length $L=17$. We can  find that no discontinuity occurs in the figures of $\mathrm{d}\omega_1/\mathrm{d}K$ and
$\mathrm{d}^2\omega_1/\mathrm{d}K^2$, which is an evidence that no phase transition has occurred. When obtaining Fig.~\ref{fig:PlotVortexDensity}(a), we can find verify that at $\gamma/\pi=0.4$, the vortex number $N_\mathrm{v}^\mathrm{G}=1$ (``Meissner phase'') and at $\gamma/\pi=0.6$, the vortex number $N_\mathrm{v}^\mathrm{G}=3$ (``vortex phase''). Now, to investigate the quasimomentum (``order parameter'') distribution of the single-particle ground state wave function $\psi_{nm}^{\left(	1\right)}$ for both cases $\gamma/\pi=0.4$  $\gamma/\pi=0.6$ and, we will calculate the Fourier transformation  $\psi^\prime_{k_xm}$ according to	
\begin{align}
	\psi'_ {k_{x}m}  =\sum_{n=-N}%
	^{N}e^{-i\left(  \gamma mn+k_{x}n\right)  }\psi_{nm}^{\left(
		1\right)  },
\end{align}
which is consistent with the transformation in Eq.~(\ref{eq:sigma-S}). And $|\psi'_{k_xm}|$ has been plotted in Figs.~\ref{fig:phase_transition}(c) and Figs.~\ref{fig:phase_transition}(d)
We can see that whether $\gamma/\pi=0.4$ [see Fig.~\ref{fig:phase_transition}(c)] or $\gamma/\pi=0.6$ []see Fig.~Fig.~\ref{fig:phase_transition}(d)] , the ``order parameter'' $k_x$ is mostly located at $k_x=0$. Thus, the ``order parameter'' does not significantly change for different ``phases''. This is another proof that no conventional phase transition has happened.   
\CIV
		 
\CIV
\subsection{Chiral current}%

\begin{figure}[ptb]
\centering\includegraphics[width=0.5\textwidth,clip]{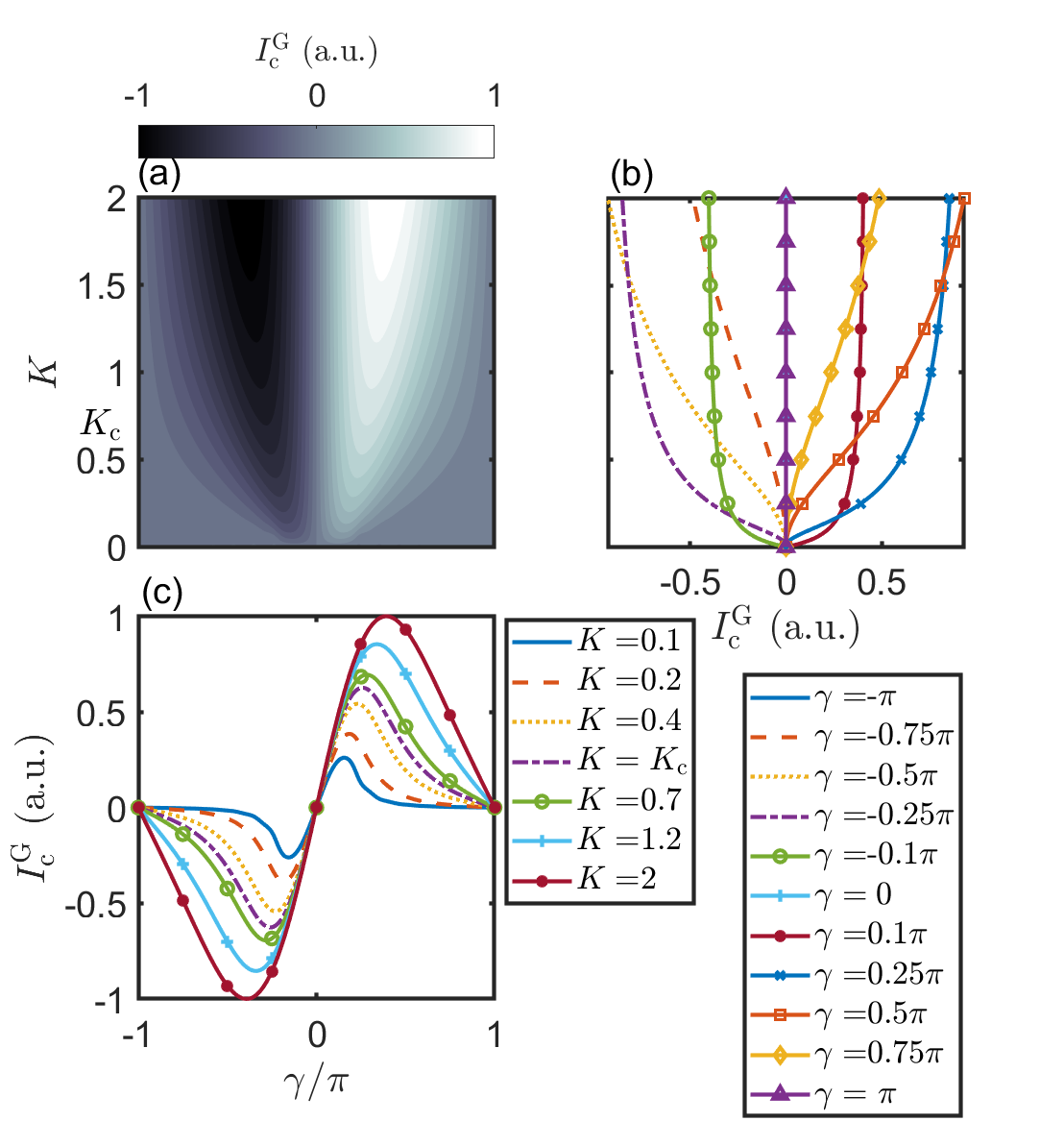}\caption{(color
online) Chiral current $I_{\text{c}}^{\text{G}}%
$ (normalized by its maximum value) plotted (a) against the
effective magnetic flux $\gamma$ and coupling ratio $K$, (b) against the
coupling ratio $K$ for different effective magnetic flux $\gamma$, and (c)
against the effective magnetic flux $\gamma$ for different coupling ratio
$K$. Note that in (c), the curves for $\gamma=0$, and $\gamma=\pm
\pi$ coincide.}
\label{fig:PlotChiralCurrent}
\end{figure}%

To better quantify the chirality manifested in the vortices, we here introduce
the chiral current operator, which, by definition, is namely
\begin{equation}
I_{\text{c}}=I_{x1}-I_{x,-1}.
\end{equation}
Here, $I_{xm}=\sum_{n}I_{nm;n+1,m}$ represents the sum of all the particle
currents along the $m$th row. The mean chiral current under the
single-particle ground state is denoted by $I_{\text{c}}^{\text{G}%
}=\left\langle G|I_{\text{c}}|G\right\rangle $, which is similar to the
definition in Eq.%
~%
(\ref{eq:ChiralCurrent}). Figure%
~%
\ref{fig:PlotChiralCurrent} has shown how the chiral current is influenced by
the effective magnetic flux $\gamma$ and the coupling ratio $K$, where
$I_{\text{c}}^{\text{G}}$ has been normalized by its maximum. In Figs.%
~%
\ref{fig:PlotChiralCurrent}(a)-(c), we see the asymmetry about the axis
$\gamma=0$ of the chiral current, an analogue to the motion of the charged
particle governed by the Lorentz force. In Fig.%
~%
\ref{fig:PlotChiralCurrent}(b), we can clearly see that when $\left\vert
\gamma\right\vert $ is increased from $0$ to $\pi$, the curve of the chiral
current against $K$ first leaves from the axis $K=0$ and then comes back. In
Fig.%
~%
\ref{fig:PlotChiralCurrent}(c), one can find that for rising $K$, the
variation of the chiral current against $\gamma$ resembles the sinusoidal
function better and better.

\section{Multi-band model}
\label{sec:multi-band model}
\begin{figure}[ptb]
\centering\includegraphics[width=0.52\textwidth,clip]{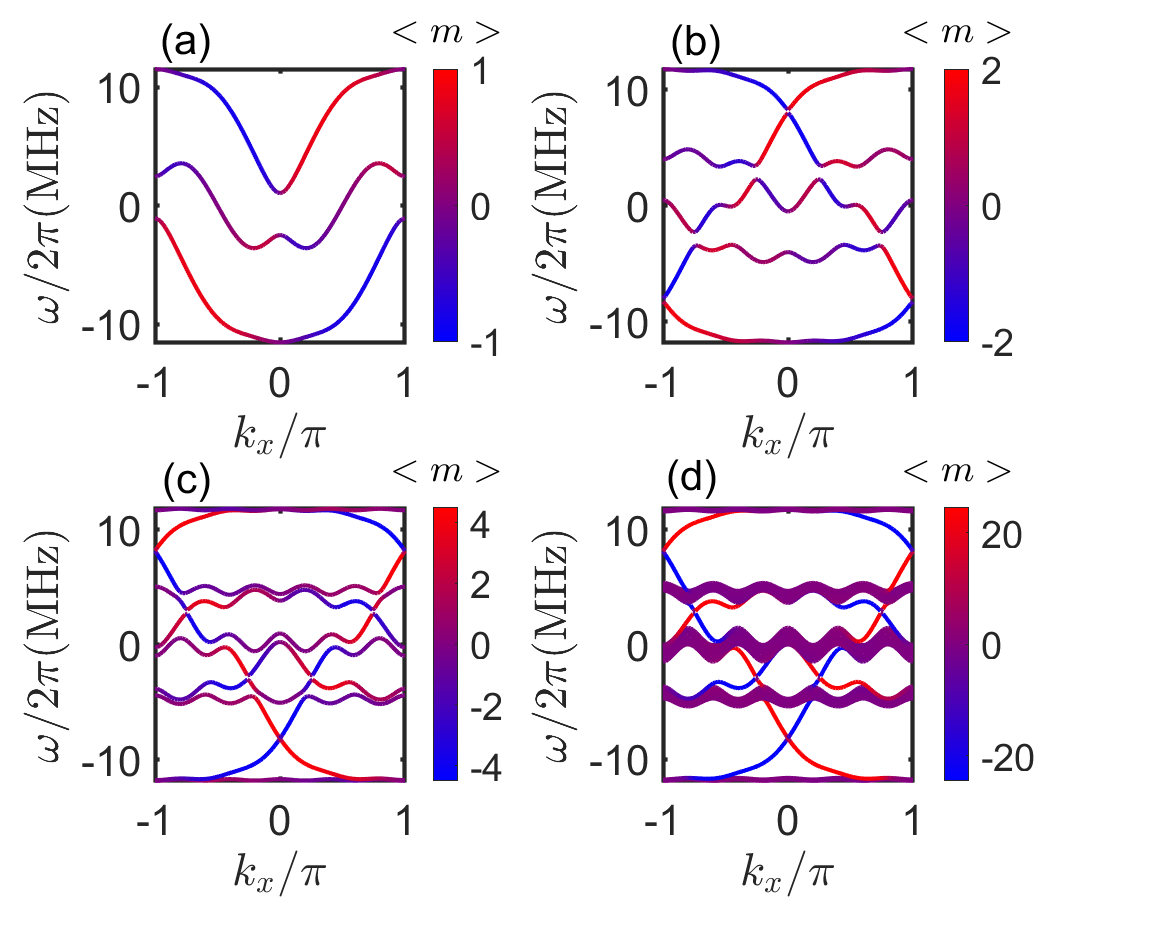}\caption{(color
online) Energy bands for the \CV lattice \CIV width (a) $W=3$, (b) $W=5$, (c) $W=10$,
and (d) $W=50$\color{black}, where $\omega$ ($k_{x}$) denotes the
eigenfrequency (row-direction quasimomentum). \color{black} We have chosen the
effective magnetic flux $\gamma=\frac{2\pi}{5}$, and coupling strengths
$g_{x}=g_{y}=2\pi\times4\operatorname{MHz}$.}%
\label{fig:E2q}%
\end{figure}

Now, we investigate how the energy bands change when the lattice width $W$
gradually becomes large, where the boundary is considered open (closed) along
the $m$ ($n$) direction. In Fig.%
~%
\ref{fig:E2q}, the energy bands are plotted for $W=50$, $10$, $5$, and $3$,
respectively, where we have chosen the effective magnetic flux $\gamma={2\pi
}/{5}$, and coupling strengths $g_{x}=g_{y}=2\pi\times4%
\operatorname{MHz}%
$. One can find that when $W$ is small [see Figs.%
~%
\ref{fig:E2q}(a) and \ref{fig:E2q}(b)], the band number is identical to $W$,
and all bands exhibit edge populations. However, when $W$ becomes large [see
Figs.%
~%
\ref{fig:E2q}(c) and \ref{fig:E2q}(d)], the band number is determined by
$\gamma$. For example, if ${\gamma}/{2\pi}$ is rational,$\,$i.e., $\gamma
=2\pi{P}/{Q}$ with $P$ and $Q$ coprime integers, the band number is namely the
denominator $Q$. That's why we see five bands in Figs.%
~%
\ref{fig:E2q}(c) and \ref{fig:E2q}(d), where $\gamma={2\pi}/{5}$ implies $P=1
$ and $Q=5$. 

The edge state in Fig.%
~%
\ref{fig:E2q} is topological phenomenon, which, according to the bulk-edge
correspondence%
~%
\cite{Hatsugai1993PRL,Hatsugai1993PRB}, can be explained via the Chern number
for periodical boundaries. In terms of the $j$th energy band, the Chern number
can be calculated via the formula%
~%
\cite{Fukui2005JPSJ}%
\begin{equation}
C_{j}=\frac{1}{2\pi i}\int\text{d}^{2}\mathbf{k}F_{xy}^{\left(  j\right)
}\left(  \mathbf{k}\right)  ,
\end{equation}
where the integrand is called the Berry curvature and defined by the bulk
eigenstate $\left\vert u_{j}\left(  \mathbf{k}\right)  \right\rangle $, i.e.,
$F_{xy}^{\left(  j\right)  }\left(  \mathbf{k}\right)  =\left\langle
\partial_{k_{x}}u_{j}\left(  \mathbf{k}\right)  \right.  \left\vert
\partial_{k_{y}}u_{j}\left(  \mathbf{k}\right)  \right\rangle -\left\langle
\partial_{k_{y}}u_{j}\left(  \mathbf{k}\right)  \right\vert \left.
\partial_{k_{x}}u_{j}\left(  \mathbf{k}\right)  \right\rangle $. The Berry
curvature reflects the $\mathbf{k}$-dependent adiabatic evolution of the bulk
eigenstate. For $\gamma=2\pi{P}/{Q}$, the unit cell possesses a size of
$1\times Q$. Thus, adopting the transformation
\begin{equation}
a_{sQ+r,m}=\frac{1}{\sqrt{WL/Q}}\sum_{\mathbf{k}}e^{ik_{y}m+isQk_{x}%
}S_{r,\mathbf{k}}^{-},
\end{equation}
where $k_{y}\in\left[  0,2\pi\right]  $ and $k_{x}\in\left[  0,2\pi/Q\right]
$, the lattice-space Hamiltonian in Eq.%
~%
(\ref{eq:H_sp}) can be transformed into $H=\sum_{\mathbf{k}}S_{r,\mathbf{k}%
}^{+}h\left(  \mathbf{k}\right)  S_{r,\mathbf{k}}^{-}$, where $S_{\mathbf{k}%
}^{-}=\left(  S_{0,\mathbf{k}}^{-},\cdots,S_{Q-1,\mathbf{k}}^{-}\right)
^{\intercal}$, $S_{r,\mathbf{k}}^{+}=\left(  S_{\mathbf{k}}^{-}\right)
^{\dag} $ and the matrix element of the single-particle Hamiltonian $h\left(
\mathbf{k}\right)  $ holds the form
\begin{align}
h_{pq}\left(  \mathbf{k}\right)  =  &  -\hbar g_{x}\left(  \delta
_{p,q+1}+\delta_{p,q-1}\right) \nonumber\\
&  -\hbar g_{x}\left[  e^{ik_{x}Q}\delta_{p,q-\left(  Q-1\right)  }%
+e^{-ik_{x}Q}\delta_{p,q+\left(  Q-1\right)  }\right] \nonumber\\
&  +2\hbar g_{y}\cos\left(  k_{y}-\gamma q\right)  \delta_{p,q}.
\end{align}
Then, via numerically diagonalizing the single-particle Hamiltonian $h\left(
\mathbf{k}\right)  $, the bulk eigenstate $\left\vert u^{\left(  j\right)
}\left(  \mathbf{k}\right)  \right\rangle $, Berry curvature $F_{xy}^{\left(
j\right)  }\left(  \mathbf{k}\right)  $, and even the Chern number can be
successively obtained.

\begin{center}%
\begin{figure*}[ptb]
\centering\includegraphics[
width=0.85\textwidth,clip
]{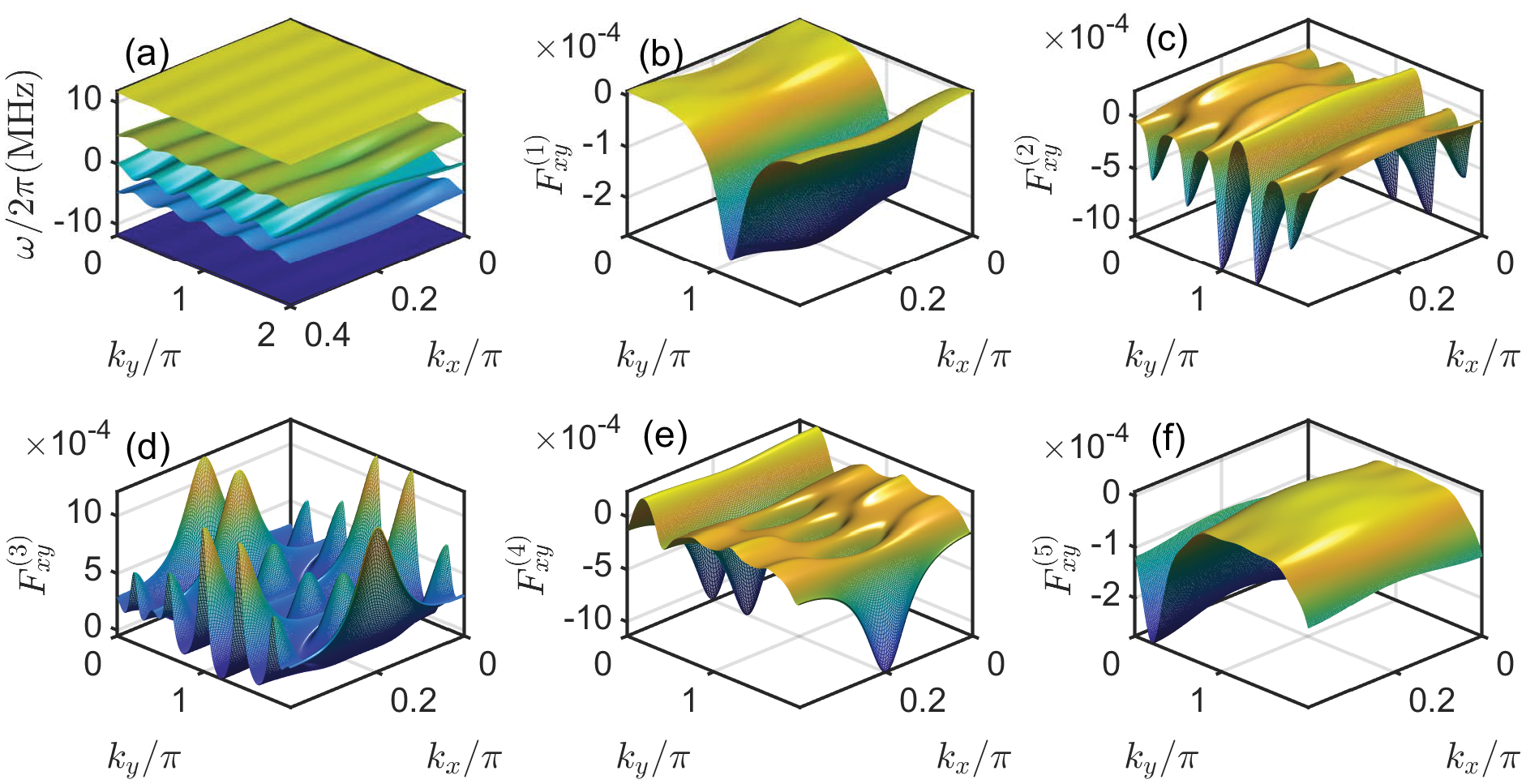}
\caption{(color online) (a) Energy bands and Berry curvatures for the
(b) first, (c) second, (d) third, (e) fourth and (f) fifth energy band\CV,
where $\omega
$ is the eigen frequency, $k_x$ ($k_y$) the row(column)-direction quasimomentum, and $F^{(j)}%
_{xy}$ the berry curvature for the $j$th energy band.
\CIV Here, we have taken the coupling strengths $g_x=g_y=2\pi\times
4~\mathrm{MHz}$
and the effective magnetic flux $\gamma=2\pi
/5$. The Chern numbers are respectively
$C_1=C_2=C_4=C_5=-1$ and $C_3=4$.}
\label{fig:BerryCurvature}
\end{figure*}%

\end{center}

In Fig.%
~%
\ref{fig:BerryCurvature}, we have focused on the special case $g_{x}%
=g_{y}=2\pi\times4%
\operatorname{MHz}%
$, where the energy bands and the Berry curvature for each band have been
plotted. The peaks or dips of the Berry curvature of mainly occur where the
bands approaches each other (see Appendix.%
~%
\ref{Sec:CalChernNum}). After integrating the Berry curvatures, the Chern
numbers can be obtained as $C_{1}=C_{2}=C_{4}=C_{5}=-1$ and $C_{3}=4$. The
winding numbers for the edge\ states in the $j$th gap can be further
calculated by the formula $I_{j}=\sum_{j^{\prime}\leq j}C_{j^{\prime}}$,
finally yielding $I_{1}=-1$, $I_{2}=-2$, $I_{3}=2$, $I_{4}=1$, and $I_{5}=0$,
which agrees with how the edge states merge and escape the bulk in Figs.%
~%
\color{black}{\ref{fig:E2q}(c) and \ref{fig:E2q}(d).}\color{black}

\section{Hofstadter butterfly}
\label{sec:HotBut}

\begin{figure}[ptb]
\centering\includegraphics[width=0.48\textwidth,clip]{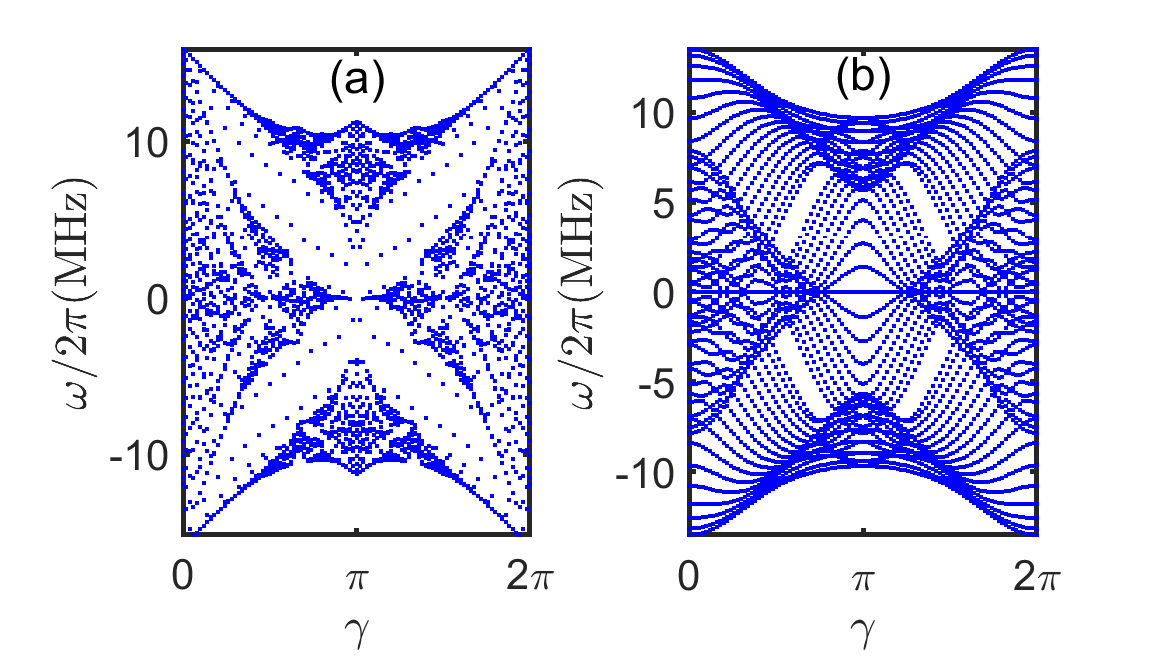}\caption{\color{black}
Eigenfrequency $\omega$ against the effective magnetic flux $\gamma$ that
changes from 0 to $2\pi$, with the lattice width $W=3$, lattice length $L=17$,
and coupling strengths $g_{x}=g_{y}=2\pi\times4\operatorname{MHz}$.
\color{black} In (a), the boundaries are open (periodical) in the \color{black}
row (column) \color{black} direction, while in (b), the boundaries are open in
both directions. }%
\label{fig:Two_But}%
\end{figure}The Harper Hamiltonian engineered can exhibit the
Hofstadter-butterfly structure%
~%
\cite{Hofstadter1976PRB} if we present the energy levels changing as the
effective magnetic flux $\gamma$. In Fig.%
~%
\ref{fig:Two_But}, we have focused on a concrete case, where the lattice width
$W=3$, lattice length $L=17$, and coupling strengths $g_{x}=g_{y}=2\pi\times4%
\operatorname{MHz}%
$. In Fig.%
~%
\ref{fig:Two_But}(a), we have chosen the $n$-open and $m$-closed boundary
condition, which reproduce the original Hofstadter problem%
~%
\cite{Hofstadter1976PRB}. However, the closed boundary condition is difficult
to implement using the current planar fabrication technology. Thus, in Fig.%
~%
\ref{fig:Two_But}(b), the $n$-open and $m$-open boundary condition is
investigated for comparison. We find the energy levels still constitute a
butterfly structure. However, the random dots in Fig.%
~%
\ref{fig:Two_But}(b) become regular in Fig.%
~%
\ref{fig:Two_But}(a). We think this can be fundamentally attributed to the $m
$-open boundary condition.

\section{Experimental details}

\label{Sec:ExpDetails}

\subsection{Generating the single-particle ground state}

\label{Sec:GenSingParticleGroundState}

To observe the chiral current patterns in the double ladder, one prerequisite
is to generate the single-particle ground state $\left\vert G\right\rangle $.
In the realistic experiment, the initial state is normally the vacuum state
$\left\vert 0\right\rangle $ after sufficient cooling in the dilution
refrigerator. Thus, we need to generate the single-particle ground sate from
the vacuum state. Here, we continue to adopt the state-generation method
employed in our previous paper on the two-leg ladder model%
~%
\cite{Zhao2020PRA}. In detail, we assume all the transmons are classically
driven, which appears as an additional term%
\begin{equation}
H_{\text{g}}=\hbar\sum_{n=-N}^{N-1}\sum_{m=-M}^{M-1}\Omega_{nm}e^{-i\nu_{m}%
t}a_{nm}^{\dag}+\text{H.c..}%
\end{equation}
in the original picture. Here, we assume the driving frequencies take $\nu
_{m}=\omega_{\text{o}}+\nu$ ($\nu_{m}=\omega_{\text{e}}+\nu$) for $m$ odd
(even). Thus, in the interaction picture, the driving Hamiltonian
$H_{\text{g}}$ becomes%
\begin{align}
H_{\text{g,I}}  &  =\hbar\sum_{n=-N}^{N-1}\sum_{m=-M}^{M-1}\Omega_{nm}e^{-i\nu
t}a_{nm}^{\dag}+\text{H.c.}\nonumber\\
&  =\hbar\sum_{j=1}^{LW}\Omega_{j}^{\prime}e^{-i\nu t}S_{j}^{\dag
}+\text{H.c.,}\label{eq:HgI}%
\end{align}
where $\Omega_{j}^{\prime}=\sum_{n=-N}^{N-1}\sum_{m=-M}^{M-1}$ $\Omega
_{nm}\psi_{nm}^{\left(  \color{black}{j} \color{black} \right)  \ast}$
represents the driving strength that stimulates the single-particle state
$\left\vert E_{j}\right\rangle $. Next, the complex driving strength
$\Omega_{nm}$ is specified as $\Omega_{nm}=\psi_{nm}^{\left(  1\right)
}\Omega$, which further transforms $H_{\text{g,I}}$ into $H_{\text{g,I}}%
=\hbar\Omega S_{1}^{\dag}e^{-i\nu t}+\mathrm{H.c..}$The detuning is further
assumed to take $\nu=E_{1}/\hbar$, and then there is an Rabi oscillation
between the vacuum sate $\left\vert 0\right\rangle $ and single-particle state
$\left\vert E_{1}\right\rangle $. If a $\frac{\pi}{2}$ pulse is applied (i.e.,
$\Omega t=\frac{\pi}{2}$), the single-particle ground state $\left\vert
G\right\rangle =\left\vert E_{1}\right\rangle $ can be obtained in just one
step from the vacuum state $\left\vert 0\right\rangle $. Suppose the
relaxation and dephasing rates are $\gamma_{nm}$ and $\Gamma_{nm}$ for the
transmon at the site $nm$, the equivalent relaxation and dephasing rates
between the states $\left\vert G\right\rangle $ and $\left\vert 0\right\rangle
$ are $\gamma_{1}=\sum_{n=-N}^{N}\sum_{m=-M}^{M}$ $\left\vert \psi
_{nm}^{\left(  1\right)  }\right\vert ^{2}\gamma_{nm}$ and $\Gamma_{1}%
=\sum_{n=-N}^{N}\sum_{m=-M}^{M}\left\vert \psi_{nm}^{\left(  1\right)
}\right\vert ^{2}\Gamma_{nm}$. Then using the Lindblad master equation, the
generation fidelity of the single-particle ground state can be calculated as
$\left\langle G\right\vert \hat{\rho}\left\vert G\right\rangle =\frac{1}%
{2}\left[  1-e^{-\frac{1}{2}\left(  \gamma_{1}+\frac{\Gamma_{1}}{2}\right)
t}\cos\left(  2\Omega t\right)  \right]  $. One can see that in the
strong-driving regime $\Omega\gg\gamma_{1},\Gamma_{1}$, the effects of the
dissipation can in principle be neglected for a $\frac{\pi}{2}$ pulse.

\subsection{Measuring the particle current}

Here, we discuss the scheme on measuring the particle current, an
indispensable step towards depicting the current pattern and then counting the
vortex number. The measurement procedure also follows our previous work%
~%
\cite{Zhao2020PRA}. For example, to measure the particle current between two
adjacent sites $nm$ and $n^{\prime}m^{\prime}$ ($n^{\prime}m^{\prime}=n+1,m$,
or $n,m+1$), we first decouple any other site that connects to $nm$ or
$n^{\prime}m^{\prime}$ and then investigate the Rabi oscillation between $nm$
and $n^{\prime}m^{\prime}$. Then, in the strong-coupling regime, $\left\vert
g_{x}\right\vert ,\left\vert g_{y}\right\vert \gg\gamma_{nm,}\Gamma_{nm}$, the
time-dependent population difference between $nm$ and $n^{\prime}m^{\prime}$
can be represented as
\begin{equation}
P_{nm;x}\left(  t\right)
\!%
=%
\!%
e^{-i\gamma_{nm;x}t}\left[  \cos%
\!%
\left(  2g_{x}t\right)  +%
\!%
\sin\left(  2g_{x}t\right)  \frac{I_{nm;x}^{\text{G}}}{g_{x}}\right]
\label{eq:PDx}%
\end{equation}
for the site index $n^{\prime}m^{\prime}=n+1,m$, where the decay factor
$\gamma_{nm;x}=\left(  \gamma_{nm}+\gamma_{n+1,m}+\Gamma_{nm}+\Gamma
_{n+1,m}\right)  /4,$ or similarly,%
\begin{equation}
P_{nm;y}\left(  t\right)
\!%
=%
\!%
e^{-i\gamma_{nm;y}t}\left[  \cos%
\!%
\left(  2g_{y}t\right)  +%
\!%
\sin\left(  2g_{y}t\right)  \frac{I_{nm;y}^{\text{G}}}{g_{y}}\right]
\label{eq:PDy}%
\end{equation}
for the site index $n^{\prime}m^{\prime}=n,m+1$, where the decay factor
$\gamma_{nm;y}=\left(  \gamma_{nm}+\gamma_{n,m+1}+\Gamma_{nm}+\Gamma
_{n,m+1}\right)  /4$. Via fitting the experimental data of the population
difference with Eqs.%
~%
(\ref{eq:PDx}) and (\ref{eq:PDy}), the particle current between the sites $nm
$ and $n^{\prime}m^{\prime}$ can in principle be extracted.

\subsection{Measuring the Hofstadter-butterfly spectrum}

The Hofstadter-butterfly spectrum can be measured using a similar method
adopted in the one-dimensional Harper model%
~%
\cite{Roushan2017Science}. From the global ground state $\left\vert
0\right\rangle =\prod_{nm}\left\vert 0\right\rangle _{nm}$, first we only
prepare the $nm$th transmon to $\left(  \left\vert 0\right\rangle
_{nm}+\left\vert 1\right\rangle _{nm}\right)  /\sqrt{2}$, where $\left\vert
1\right\rangle _{nm}=\sigma_{nm}^{+}\left\vert 0\right\rangle _{nm}$ and
$\sigma_{nm}^{+}=\left\vert 1\right\rangle _{nm}\left\langle 0\right\vert
_{nm}$ denotes the single-qubit raising operator. Here, note that in the
single-particle regime, $\sigma^{+}_{nm}$ ($\sigma^{-}_{nm}$) is equivalent to
the bosonic creation (annihilation) operator $a^{\dagger}_{nm}$ ($a_{nm}$).
Then, the product state of all the qubits takes $\left(  \left\vert
0\right\rangle +\left\vert 1_{nm}\right\rangle \right)  /\sqrt{2}$ with
$\left\vert 1_{nm}\right\rangle =\sigma_{nm}^{+}\left\vert 0\right\rangle $.
Then, we synthesize the specific two-dimensional Harper Hamiltonian [see Eq.%
~%
(\ref{eq:H_sp})] and record the evolution of the raising operator, termed by
$\chi_{nm}\left(  t\right)  =2\left\langle \sigma_{nm}^{+}\left(  t\right)
\right\rangle $. Technically, the quantity $\chi_{nm}\left(  t\right)  $
amounts to measurements of the Pauli X operator $\sigma_{nm}^{x}$ and Pauli Y
operator $\sigma_{nm}^{y}$ considering $\sigma_{nm}^{+}=$\ $\left(
\sigma_{nm}^{x}+i\sigma_{nm}^{y}\right)  /2$, finally leading to $\chi
_{nm}\left(  t\right)  =\left\langle \sigma_{nm}^{x}\left(  t\right)
\right\rangle +i\left\langle \sigma_{nm}^{y}\left(  t\right)  \right\rangle
$\textrm{. }When we vary $nm$ to cover all the sites, the initial states
construct a complete basis and thus each energy eigenstate will have some
overlap with at least one initial state, which guarantees that the energy
spectrum is fully resolved. Thus, we will average $\chi_{nm}\left(  t\right)
$ over all the sites and obtain $\bar{\chi}\left(  t\right)  =\frac{1}{LW}%
\sum_{n=-N}^{N}\sum_{m=-M}^{M}\chi_{nm}\left(  t\right)  $. The strict
analysis in Appendix.%
~%
\ref{Append:Hof_measurement} presents that the frequencies that occur in the
Fourier transform of $\chi\left(  t\right)  $, with a homogeneous amplitude
$\frac{1}{LW}$, are the eigenenergies of the Harper Hamiltonian.

\subsection{Measuring the energy bands: a special method\label{sec:MEB}}

We now discuss how to measure the energy bands in Fig.%
~%
\ref{fig:E2q}. The detailed method is based on making Fourier transforms of
single-particle energy eigenstates in the open-boundary conditions of both
directions. We still denote the single-particle energy eigenstate by
$\left\vert E_{j}\right\rangle =\sum_{nm}\psi_{nm}^{\left(  j\right)
}\left\vert 1_{nm}\right\rangle =\sum_{nm}\psi_{nm}^{\left(  j\right)  }%
\sigma_{nm}^{\dag}\left\vert 0\right\rangle $ for the eigen energy $E_{j}$
(see Sec.%
~%
\ref{Sec:CurrentPattern}).

First, we should synthesize the single-particle state $\left\vert
E_{j}\right\rangle $. The method is identical to the generation of
single-particle ground state (see Sec.%
~%
\ref{Sec:GenSingParticleGroundState}) except that $\psi_{nm}^{\left(
1\right)  }$ is replaced with $\psi_{nm}^{\left(  j\right)  }$ (e.g.,
$\Omega_{nm}=\psi_{nm}^{\left(  j\right)  }\Omega$).

Then, we investigate how to measure the state $\left\vert E_{j}\right\rangle
$, or rather, $\psi_{nm}^{\left(  j\right)  }$. Although the actual state
could vary due to experimental errors, we still use the same symbols here for
convenience. The direct measurement of all the transmons will provide
$\left\vert \psi_{nm}^{\left(  j\right)  }\right\vert ^{2}$ or equivalently,
$\left\vert \psi_{nm}^{\left(  j\right)  }\right\vert $.

What remains to be determined is the phase of $\psi_{nm}^{\left(  j\right)  }$
, which we denote by $\theta_{nm}^{\left(  j\right)  }$. To do this, the
strategy is to measure the relative phase between adjacent sites, e.g.,
$\theta_{nm}^{\left(  j\right)  }-\theta_{n^{\prime}m^{\prime}}^{\left(
j\right)  }$ between the sites $nm$ and $n^{\prime}m^{\prime}$ [$n^{\prime
}m^{\prime}=(n+1,m)$ or $n^{\prime}m^{\prime}=(n,m+1)$]. This pair of sites is
immediately decoupled from the other sites once $\left\vert E_{j}\right\rangle
$ is generated. Meanwhile, the intersite coupling strength $g_{nm;n^{\prime
}m^{\prime}}$ is tuned appropriately for a time \CV$\tau_{1}$ \CIV to generate what we
call a X-$\frac{\pi}{4}$ pulse ($g_{nm;n^{\prime}m^{\prime}}\CV\tau_{1}\CIV=\frac{\pi
}{4}$), which gives the final state $\psi_{nm}^{\left(  j\right)  }\left(
t_{1}\right)  =\left[  \psi_{nm}^{\left(  j\right)  }-i\psi_{n^{\prime
}m^{\prime}}^{\left(  j\right)  }\right]  /\sqrt{2}$ and $\psi_{n^{\prime
}m^{\prime}}^{\left(  j\right)  }\left(  t_{1}\right)  =\left[  \psi
_{n^{\prime}m^{\prime}}^{\left(  j\right)  }-i\psi_{nm}^{\left(  j\right)
}\right]  /\sqrt{2}$ \CV with $t_1=\tau_1$\CIV. Now, the measurement of the qubits at both sites
provides the experimental data of $\left\vert \psi_{nm}^{\left(  j\right)
}\left(  t_{1}\right)  \right\vert ^{2}$ and $\left\vert \psi_{n^{\prime
}m^{\prime}}^{\left(  j\right)  }\left(  t_{1}\right)  \right\vert ^{2}$,
either of which will offer the same information of the relative phase between
both sites. For example, the datum of $\left\vert \psi_{nm}^{\left(  j\right)
}\left(  t_{1}\right)  \right\vert ^{2}$ gives the relation
\begin{align}
\left\vert \psi_{nm}^{\left(  j\right)  }\left(  t_{1}\right)  \right\vert
^{2}  &  =\frac{1}{2}\left\vert \psi_{nm}^{\left(  j\right)  }\right\vert
^{2}+\frac{1}{2}\left\vert \psi_{n^{\prime}m^{\prime}}^{\left(  j\right)
}\right\vert ^{2}\nonumber\\
&  +\left\vert \psi_{nm}^{\left(  j\right)  }\right\vert \left\vert
\psi_{n^{\prime}m^{\prime}}^{\left(  j\right)  }\right\vert \sin\left[
\theta_{n^{\prime}m^{\prime}}^{\left(  j\right)  }-\theta_{nm}^{\left(
j\right)  }\right]  .\label{eq:angle_diff_sin}%
\end{align}
This implies $\sin\left[  \theta_{nm}^{\left(  j\right)  }-\theta_{n^{\prime
}m^{\prime}}^{\left(  j\right)  }\right]  $ can be extracted as $\left\vert
\psi_{nm}^{\left(  j\right)  }\right\vert $ and $\left\vert \psi_{n^{\prime
}m^{\prime}}^{\left(  j\right)  }\right\vert $ have been known. However, to
uniquely determine $\theta_{nm}^{\left(  j\right)  }-\theta_{n^{\prime
}m^{\prime}}^{\left(  j\right)  }$, we still need the information of
$\cos\left[  \theta_{nm}^{\left(  j\right)  }-\theta_{n^{\prime}m^{\prime}%
}^{\left(  j\right)  }\right]  $. This can be performed via inserting a free
rotation of both qubits before applying the X-$\frac{\pi}{4}$ pulse. To
realize this, we switch off the coupling between both sites and \CV adiabatically \CIV detune the
qubit frequency $\omega_{nm}$ ($\omega_{n^{\prime}m^{\prime}}$) with the shift
$\frac{\Delta_{nm}}{2}$ ($-\frac{\Delta_{nm}}{2}$). After a time \CV $\tau_{1}$\CIV, we
generate what we call a Z-$\frac{\pi}{2}$ pulse ($\Delta_{nm}\CV\tau_{1}\CIV=\frac
{\pi}{2}$). Then, \CV we adiabatically tune the frequency shift $\frac{\Delta_{nm}}{2}$ ($-\frac{\Delta_{nm}}{2}$) back to zero, and thus \CIV the state components $\psi_{nm}^{\left(  j\right)  }$ and
$\psi_{n^{\prime}m^{\prime}}^{\left(  j\right)  }$ evolve to $\psi
_{nm}^{\left(  j\right)  }\left(  \CV t_{1}\CIV\right)  =e^{\frac{-i\pi}{4}}\psi
_{nm}^{\left(  j\right)  }$ and $\psi_{n^{\prime}m^{\prime}%
}^{\left(  j\right)  }\left(  \CV t_{1}\CIV\right)  =e^{\frac{i\pi}{4}}\psi
_{n^{\prime}m^{\prime}}^{\left(  j\right)  }$ \CV with $t_1=\tau_1$\CIV. Next, we apply
the X-$\frac{\pi}{4}$ pulse ($g_{nm;n^{\prime}m^{\prime}}\CV \tau_{2}\CIV=\frac{\pi}{4}%
$) and the state components further evolve to $\psi_{nm}^{\left(  j\right)
}\left(\CV t_{2}\CIV\right)  =\left[  \psi_{nm}^{\left(  j\right)  }\left(
\CV t_{1}\CIV\right)  -i\psi_{n^{\prime}m^{\prime}}^{\left(  j\right)  }\left(
\CV t_{1}\CIV\right)  \right]  /\sqrt{2}$ and $\psi_{n^{\prime}m^{\prime}}^{\left(
j\right)  }\left( \CV t_{2}\CIV\right)  =\left[  \psi_{n^{\prime}m^{\prime}%
}^{\left(  j\right)  }\left( \CV t_{1}\CIV\right)  -i\psi_{nm}^{\left(  j\right)
}\left( \CV t_{1}\CIV\right)  \right]  /\sqrt{2}$ \CV with $t_2=t_1+\tau_2$\CIV. Afterwards, the readout of both
qubits is implemented, and thus both $\left\vert \psi_{nm}^{\left(  j\right)
}\left( \CV t_2\CIV\right)  \right\vert ^{2}$ and $\left\vert \psi
_{n^{\prime}m^{\prime}}^{\left(  j\right)  }\left( \CV t_{2}\CIV\right)  \right\vert
^{2}$ are accessible quantities. The measurement of the $nm$th qubit gives the
relation%
\begin{align}
\left\vert \psi_{nm}^{\left(  j\right)  }\left( \CV t_2\CIV\right)
\right\vert ^{2}  &  =\frac{1}{2}\left\vert \psi_{nm}^{\left(  j\right)
}\right\vert ^{2}+\frac{1}{2}\left\vert \psi_{n^{\prime}m^{\prime}}^{\left(
j\right)  }\right\vert ^{2}\nonumber\\
&  +\left\vert \psi_{nm}^{\left(  j\right)  }\right\vert \left\vert
\psi_{n^{\prime}m^{\prime}}^{\left(  j\right)  }\right\vert \cos\left[
\theta_{n^{\prime}m^{\prime}}^{\left(  j\right)  }-\theta_{nm}^{\left(
j\right)  }\right] \label{eq:angle_diff_cos}%
\end{align}
From Eqs.%
~%
(\ref{eq:angle_diff_sin}) and (\ref{eq:angle_diff_cos}), the relative phase
$\theta_{n^{\prime}m^{\prime}}^{\left(  j\right)  }-\theta_{nm}^{\left(
j\right)  }$ can be uniquely determined if it is confined to the regime
$\left(  -\pi,\pi\right]  $. Now, the measurement of the relative phase
between to adjacent sites is completed. To determine the phases of all wave
function components, we should measure all the relative phases between the
sites $nm$ and $n+1,m$ for $-N\leq n\leq N-1$ and $-M\leq m\leq M$, and those
between $-N,m$ and $-N,m+1$ for $-M\leq m\leq M-1$. Supposing $\psi
_{-N,\_M}\geq0$, then the phase of $\psi_{nm}^{\left(  j\right)  }$ can be
represented as $\theta_{nm}^{\left(  j\right)  }=$ $\sum_{p=-M+1}^{m}\left[
\theta_{-N,p}^{\left(  j\right)  }-\theta_{-N,p-1}^{\left(  j\right)
}\right]  +\sum_{p=-N+1}^{n}\left[  \theta_{p,m}^{\left(  j\right)  }%
-\theta_{p-1,m}^{\left(  j\right)  }\right]  .$

\begin{figure}[ptb]
\centering\includegraphics[width=0.52\textwidth,clip]{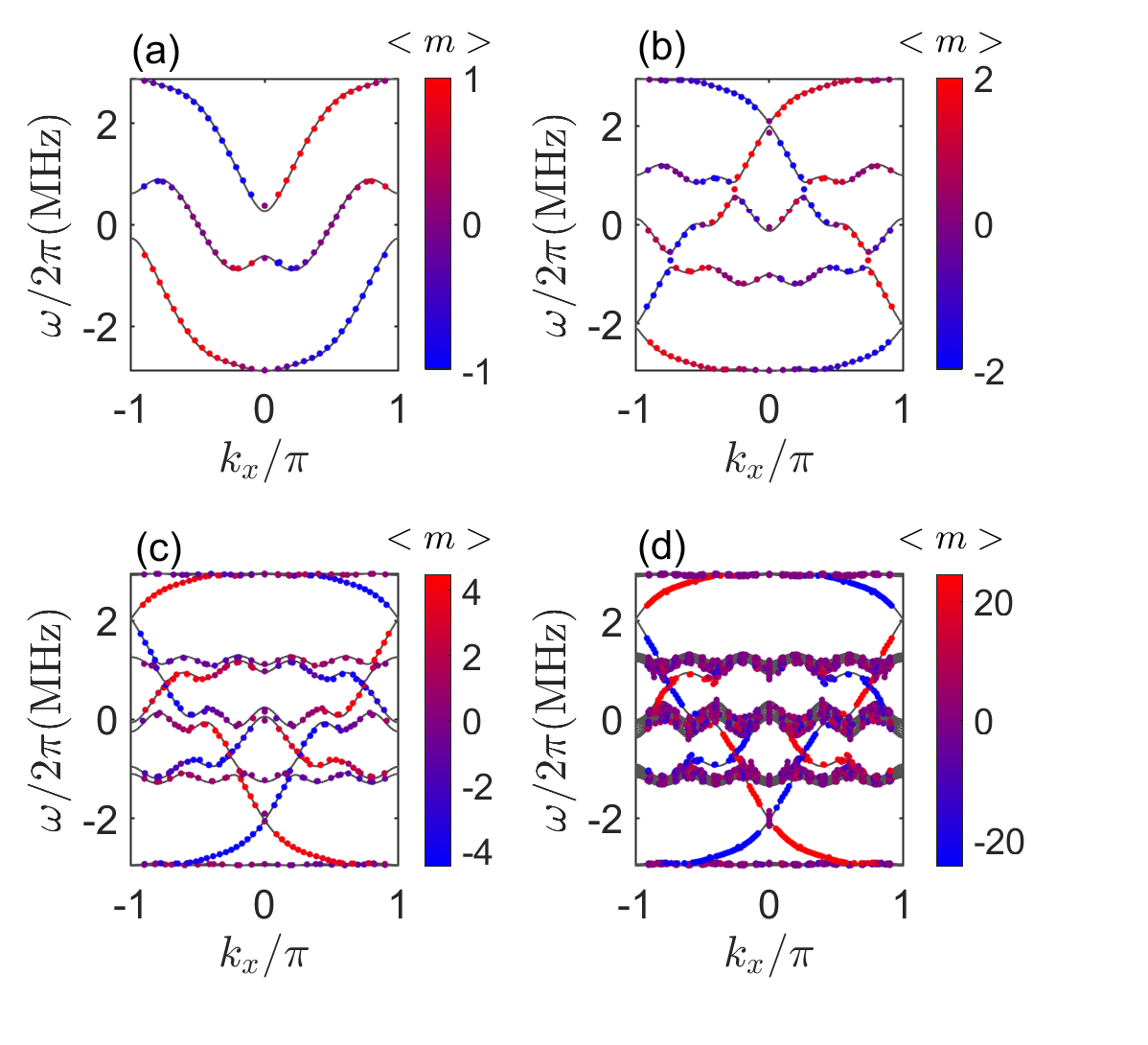}\caption{(color
online) Measured energy bands (colored dots) for the lattice width (a) $W=3$,
(b) $W=5$, (c) $W=10$, and (d) $W=50$. \color{black} Here, $\omega$ ($k_{x}$)
is the eigenfrequency (row-direction quasimomentum). Besides, we \color{black}
have chosen the effective magnetic flux $\gamma=\frac{2\pi}{5}$, and coupling
strengths $g_{x}=g_{y}=2\pi\times4\operatorname{MHz}$. The grey solid curves
are strict energy bands (see Fig.~\ref{fig:E2q}) for comparison.}%
\label{fig:randpoints}%
\end{figure}

The last step for constructing the energy bands is to make space Fourier
transformation of $\psi_{nm}^{\left(  j\right)  }$. The single-particle state
$\psi_{nm}^{\left(  j\right)  }$ must be the superposition of the states with
the same energy in Fig.%
~%
\ref{fig:E2q}, that is,%
\begin{equation}
\left\vert E_{j}\right\rangle =\sum_{n=-N}^{N}\sum_{m=-M}^{M}\psi
_{nm}^{\left(  j\right)  }a_{nm}^{\dag}\left\vert 0\right\rangle =\sum
_{\hbar\omega\left(  k_{x}\right)  =E_{j}}\CV\psi_{k_{x}m}'\CIV b_{k_{x},m}^{\dag
}\left\vert 0\right\rangle .
\end{equation}
From Eq.%
~%
(\ref{eq:sigma-S}), which establishes the relation between $a_{nm}$ and
$b_{k_{x},m}$, we can further obtain the Fourier expansion of $\psi
_{nm}^{\left(  j\right)  }$, which is
\begin{equation}
\psi_{nm}^{\left(  j\right)  }=\sum_{\hbar\omega\left(  k_{x}\right)  =E_{j}%
}\frac{\CV\psi_{k_{x}m}'\CIV}{\sqrt{L}}e^{i\left(  \gamma m+k_{x}\right)  n}.
\end{equation}
To quantify the quasimomentum distribution of $\psi_{nm}^{\left(  j\right)  }%
$, we calculate the Fourier transformation of $\psi_{nm}^{\left(  j\right)  }$
and seek their square sum over $m$, thus leading to the quantity
\begin{align}
\CV P^{(j)}\CIV\left(  k_{x}\right)   &  =\sum_{m=-M}^{M}\left\vert \sum_{n=-N}%
^{N}\frac{e^{-i\left(  \gamma mn+k_{x}n\right)  }}{\sqrt{L}}\psi_{nm}^{\left(
j\right)  }\right\vert ^{2}\nonumber\\
&  =\sum_{m=-M}^{M}\left\vert \sum_{\hbar\omega\left(  k_{x}^{\prime}\right)
=E_{j}}\frac{\psi_{k_{x}^{\prime}m}}{L}\frac{\sin\frac{\left(  k_{x}%
-k_{x}^{\prime}\right)  L}{2}}{\sin\frac{k_{x}-k_{x}^{\prime}}{2}}\right\vert
^{2}.\label{eq:PIj}%
\end{align}
Via seeking the peak value of $\CV P^{(j)}\CIV\left(  k_{x}\right)  $, we obtain the
point ($k_{x}^{\prime}$, $\omega\left(  k_{x}^{\prime}\right)  $) where
$\hbar\omega\left(  k_{x}^{\prime}\right)  =E_{j}$. When we plot all the
points ($k_{x}^{\prime},\omega\left(  k_{x}^{\prime}\right)  )$, the energy
bands can in principle be restored. In Fig.%
~%
\ref{fig:randpoints}, the measured energy bands are shown by the colored dots.
One can see that they agree very well with the strict results indicated by the
solid grey curves (see Fig.%
~%
\ref{fig:E2q}). Via observing how the edge states emerge into and escape from
the bulk states, we can then obtain the winding number and also the Chern
number%
~%
\cite{Hatsugai1993PRB,Hatsugai1993PRL}.

However, the method introduced here needs many details of the Hamiltonian. For
example, when generating the energy eigenstate $\left\vert E_{j}\right\rangle
$, we need its wave function $\psi_{nm}^{\left(  j\right)  } $, which is
fundamentally determined by the concrete Hamiltonian parameters. The
contradiction lies in that if we know all the details of the Hamiltonian, we
can straightforwardly calculate the energy bands. The significance of
measuring the energy eigenstate $\left\vert E_{j}\right\rangle $ and also
$\omega$-$k_{x}$ dependence using this method is mostly exhibited in testing
whether the experimental results agree with the theoretical ones. Thus, in the
next subsection Sec.%
~%
\ref{Append:GMMEB}, we will develop a method of measuring energy bands without
knowing all the details of the Hamiltonian. But the latter does pose a way
more severe requirement on the decoherence time if the qubit number is very large.

\subsection{Measuring the energy bands: a general method}

\label{Append:GMMEB}%

\CV
Here, we will introduce a more general method to measure the energy bands. By
exciting the qubit at $nm$ from the global ground state $\left\vert
0\right\rangle $, we can first generate the initial system state
\begin{align}
\left\vert \psi_{nm}\left(  0\right)  \right\rangle  &  =\frac{1}{\sqrt{2}%
}\left\vert 1_{nm}\right\rangle +\frac{1}{\sqrt{2}}\left\vert 0\right\rangle
\nonumber\\
&  =\frac{1}{\sqrt{2}}\sum_{j=1}^{LW}\psi_{nm}^{\left(  j\right)  \ast
}\left\vert E_{j}\right\rangle +\frac{1}{\sqrt{2}}\left\vert 0\right\rangle ,
\end{align}
where the relation $\left\vert 1_{nm}\right\rangle =a_{nm}^{\dag}\left\vert
0\right\rangle =\sum_{j=1}^{LW}\psi_{nm}^{\left(  j\right)  \ast}S_{j}^{\dag
}\left\vert 0\right\rangle =\sum_{j=1}^{LW}\psi_{nm}^{\left(  j\right)  \ast
}\left\vert E_{j}\right\rangle $ [see Eq.%
~%
(\ref{eq:Sigma_j_p})] has been used. Immediately after that, we engineer the
Harper Hamiltonian $H_{\text{I}}$ (see Eq.%
~%
\ref{eq:H_sp}) and wait for some time $t$, until the system further evolves
to
\begin{equation}
\left\vert \psi_{nm}\left(  t\right)  \right\rangle =\frac{1}{\sqrt{2}}%
\sum_{j=1}^{LW}\psi_{nm}^{\left(  j\right)  \ast}e^{-i\omega_{j}t}\left\vert
E_{j}\right\rangle +\frac{1}{\sqrt{2}}\left\vert 0\right\rangle
\label{eq:psi_general}%
\end{equation}
with the single-particle eigen frequency $\omega_{j}=E_{j}/\hbar$. 

Now, we will show that the wave function $\left\vert \psi_{nm}\left(  t\right)  \right\rangle $
in Eq.%
~%
(\ref{eq:psi_general})\ can be measured using a similar method mentioned in
Sec.%
~%
\ref{sec:MEB}. For convenience of description, we first represent $|\psi_{nm}(t)\rangle$ in the qubit basis as 
\begin{equation}
	\left\vert \psi_{nm}\left(  t\right)  \right\rangle =
	\sum_{\substack{p=-N, q=-M}}^{\substack{N,M}} \psi_{nm;pq}(t)\left\vert
	1_{pq}\right\rangle +\psi_{nm;0}(t)\left\vert 0\right\rangle
	\label{eq:psi_general_qubit}.
\end{equation}
To uniquely determine the state $|\psi_{nm}\rangle$, we need to know all the wave function components $\psi_{nm;pq}(t)$ and $\psi_{nm;0}(t)$. Via multiple-qubit measurement, we can straightforwardly obtain the populations on $\left\vert
1_{nm}\right\rangle$ and $\vert 0\rangle$, that is, $|\psi_{nm;pq}(t)|^2$ and $|\psi_{nm;0}(t)|^2$. Equivalently, it means we can know $|\psi_{nm;pq}(t)|$ and $|\psi_{nm;0}(t)|$ directly. All remains to be done is to determine the phase angle of all the wave function amplitudes. Not losing any generality, we assume the convenient wave function gauge $\arg\{\psi_{nm;0}\}=0$, which, otherwise, will only add a global phase factor  $\arg\{\psi_{nm;0}\}=0$ to the whole state vector $|\psi_{nm}(t)\rangle$ where this global phase factor must be subtracted to guarantee the validity of Eq.~(\ref{eq:psi_tilde_w}) in the following. Then, once $\arg\{\psi_{nm;pq}\}$ is obtained, one can finally determine the wave function $|\psi_{nm}(t)\rangle$.

Next, we focus on how to determine $\theta_{nm;pq}\left(  t\right)  =\arg\left\{  \psi_{nm;pq}\left(  t\right)  \right\}  $, the phase angle of $\psi_{nm;pq}$. To do this, we can first use what we call the carrier
process. In detail, we switch off all the inter-qubit couplings and, meanwhile,
resonantly drive the qubit Q$_{pq}$ with the
driving strength $\Omega_{pq}/2$. After what we call a
X-$\frac{\pi}{2}$ pulse ($\Omega_{pq}\tau_{1}=\frac{\pi}{2}$), we
denote the final state as
\begin{align}
\left\vert \psi_{nm}(t_1)\right\rangle    =&	\sum_{\substack{p'=-N, q'=-M}}^{\substack{N,M}} {\psi_{nm;p'q'}(t_1)}
\left\vert 1_{p'q'}\right\rangle \nonumber \\
&+	\sum_{\substack{p'=-N, q'=-M\\p'q'\neq pq}}^{\substack{N,M}}
\psi'_{nm;p'q'}(t_1)\left\vert 1_{p'q'%
}1_{pq}\right\rangle\nonumber \\
&+\psi_{nm;0}(t_1)\left\vert 0\right\rangle ,
\end{align}
where $t_1=t+\tau_1$,
$\left\vert 1_{p'q'}1_{pq}\right\rangle =a_{p'q'}^{\dag}a_{pq}^{\dag}\left\vert 0\right\rangle $ is the
double-particle state, and $\psi'_{nm;p'q'}(t_1)$ denotes the wave function component on the state $\left\vert 1_{p'q'}1_{pq}\right\rangle$. Due to the Rabi oscillation between  $\left\vert 0\right\rangle $ and $\left\vert
1_{pq}\right\rangle $, there
should be the relations
\begin{align}
\psi_{nm;0}(t_1)  &  =\frac{1}{\sqrt{2}}\left[  \psi_{nm;0}\left(  t\right)  -i\psi_{nm;pq}\left(
t\right)  \right],\\
\psi_{nm;pq}(t_1)  &  =\frac{1}{\sqrt{2}}\left[
\psi_{nm;pq}\left(  t\right)  -i\psi_{nm;0}\left(
t\right)  \right]  ,
\end{align}
while the Rabi oscillation between $\left\vert 1_{p'q'}\right\rangle $ and
$\left\vert 1_{p'q'}1_{pq}\right\rangle $ ($p'q'\neq pq$) gives $\psi_{nm;p'q'}(t_1)=\psi_{nm;p'q'}\left(  t\right)
/\sqrt{2} $ and $\psi'_{nm;p'q'}(t_1)=-i\psi
_{nm;p'q'}\left(  t\right)  /\sqrt{2}$, which are however out of our interest.
Now, measuring the qubits will give the probability on $\left\vert
1_{pq}\right\rangle $, i.e.,
\begin{equation}
\left\vert \psi_{nm;pq}(t_1)\right\vert ^{2}=\frac
{\left\vert \psi_{nm;pq}\left(  t\right)  \right\vert ^{2}}%
{2}+\frac{\left\vert \psi_{nm;0}\left(  t\right)  \right\vert ^{2}}%
{2}-\sin\theta_{nm;pq}\left(  t\right)  ,\label{eq:chi_pq_sin}%
\end{equation}
where we mention again the notation $\theta_{nm;pq}\left(  t\right)  =\arg\left\{  \psi
_{nm;pq}\left(  t\right)  \right\}  $, and also $\arg\left\{  \psi_{nm;0%
}\left(  t\right)  \right\}  =0$ has already been assumed. To uniquely determine
$\theta_{nm;pq}\left(  t\right)  $, we also need know
$\cos\theta_{nm;pq}\left(  t\right)  $. To do this, a free
rotation of the qubit at $pq$ should be inserted before
applying the X-$\frac{\pi}{2}$ pulse. This can be realised by switching off
the driving field ($\Omega_{pq}=0$) but 
adiabatically detune the qubit Q$_{pq}$ with a frequency shift $\Delta_{pq}$. After a
time $\tau_{1}$, we generate what we call a Z-$\frac{\pi}{2}$ pulse
($\Delta_{p'q'}\tau_{1}=\frac{\pi}{2}$). Then,  we tune the frequency shift $\Delta_{pq}$ back to zero and thus, the state component
$\psi_{nm;pq}\left(  t\right)  $ will evolve to $-i\psi
_{nm;pq}\left(  t\right)  $ but other components are
unchanged. Next, we also drive the qubit Q$_{pq}$ with the strength
$\Omega_{pq}/2$ that forms a X-$\frac{\pi}{2}$ pulse
($\Omega_{pq}\tau_{2}=\frac{\pi}{2}$). Now, we denote the final
state as
\begin{align}
	\left\vert \psi_{nm}(t_2)\right\rangle    =&	\sum_{\substack{p'=-N, q'=-M}}^{\substack{N,M}} {\psi_{nm;p'q'}(t_2)}
	\left\vert 1_{p'q'}\right\rangle \nonumber \\
	&+	\sum_{\substack{p'=-N, q'=-M\\p'q'\neq pq}}^{\substack{N,M}}
	\psi'_{nm;p'q'}(t_2)\left\vert 1_{p'q'%
	}1_{pq}\right\rangle\nonumber \\
	&+\psi_{nm;0}(t_2)\left\vert 0\right\rangle ,
\end{align}
where $t_2=t+\tau_1+\tau_2$.
The Rabi oscillation between  $\left\vert 0\right\rangle $
and $\left\vert 1_{pq}\right\rangle $ will render the relations
\begin{align}
\psi_{nm;0}(t_2)   &  =\frac{1}{\sqrt{2}}\left[  \psi_{nm;0}\left(  t\right)  -\psi_{nm;pq}(t)  
\right]  ,\\
\psi_{nm;pq}(t_2)  &  =\frac{-i}{\sqrt{2}}\left[
\psi_{nm;pq}(t)   +\psi_{nm;0}\left(
t\right)  \right]  ,
\end{align}
while the Rabi oscillation between $\left\vert 1_{p'q'}\right\rangle $ and
$\left\vert 1_{p'q'}1_{pq}\right\rangle $ ($p'q'\neq pq$) gives $\psi_{nm;p'q'}(t_2)=\psi_{nm;p'q'}\left(  t\right)
/\sqrt{2}$ and $\psi'_{nm;p'q'}(t_2)=-i\psi_{nm;p'q'}\left(
t\right)  /\sqrt{2}$, which are however out of our interest. Now measuring the
qubits will give the probability on $\left\vert 1_{pq%
}\right\rangle $
\begin{align}
\left\vert \psi_{nm;pq}(t_2)\right\vert ^{2}  &  =\frac{1}%
{2}\left\vert \psi_{nm;pq}\left(  t\right)  \right\vert
^{2}+\frac{1}{2}\left\vert \psi_{nm;0}\left(  t\right)  \right\vert
^{2}\nonumber\\
&  +\left\vert \psi_{nm;pq}\left(  t\right)  \right\vert
\left\vert \psi_{nm;0}\left(  t\right)  \right\vert \cos\theta
_{nm;pq}\left(  t\right)  .\label{eq:chi_pq_cos}%
\end{align}
Here, Eqs.%
~%
(\ref{eq:chi_pq_sin}) and (\ref{eq:chi_pq_cos}) will together give the phase
angle $\theta_{nm;pq}\left(  t\right)  $. Since the magnitude
of $\psi_{nm;pq}\left(  t\right)  $ and its phase angle $\arg\{\psi_{nm;pq}\left(  t\right) \}=\theta_{nm;pq}\left(  t\right)=\arg\{\cos\theta_{nm;pq}\left(  t\right)+i \sin\theta_{nm;pq}\left(  t\right) \}$
can be obtained for all possible $pq$, we can hence reconstruct the wave function $\left\vert
\psi_{nm}\left(  t\right)  \right\rangle $ through the amplitudes $\psi_{nm;pq}\left(  t\right)  $ and $\psi_{nm;0}\left(  t\right)  $.
\CIV

To find all the eigen frequencies $\omega_{j}$, we first perform the
time-domain Fourier transformation of $\left\vert \psi_{nm}\left(  t\right)
\right\rangle $ [see Eq.%
~%
(\ref{eq:psi_general})], giving
\begin{align}
\left\vert \tilde{\psi}_{nm}\left(  \omega\right)  \right\rangle =  &
\frac{1}{T}\int_{0}^{T}e^{i\omega t}\left\vert \psi_{nm}\left(  t\right)
\right\rangle \text{d}t\nonumber\\
=  &  \frac{1}{\sqrt{2}}\sum_{j}\psi_{nm}^{\left(  j\right)  \ast}%
e^{i\frac{\left(  \omega-\omega_{j}\right)  T}{2}}\frac{\sin\frac{\left(
\omega-\omega_{j}\right)  T}{2}}{\frac{\left(  \omega-\omega_{j}\right)  T}%
{2}}\left\vert E_{j}\right\rangle \nonumber\\
&  +\frac{1}{\sqrt{2}}e^{i\frac{\omega T}{2}}\frac{\sin\frac{\omega T}{2}%
}{\frac{\omega T}{2}}\label{eq:psi_tilde_w}\text{.}%
\end{align}
Please note that $\left\vert \tilde{\psi}_{nm}\left(  \omega\right)
\right\rangle $ is not necessarily normalized. Next, we define what we call
the feature function
\begin{align}
F\left(  \omega\right)   &  =\sum_{nm}\left\langle \tilde{\psi}_{nm}\left(
\omega\right)  \right.  \left\vert \tilde{\psi}_{nm}\left(  \omega\right)
\right\rangle \nonumber\\
&  =\frac{1}{2}\sum_{j}\frac{\sin^{2}\frac{\left(  \omega-\omega_{j}\right)
T}{2}}{\left[  \frac{\left(  \omega-\omega_{j}\right)  T}{2}\right]  ^{2}%
}+\frac{1}{2}\frac{\sin^{2}\frac{\omega T}{2}}{\left(  \frac{\omega T}%
{2}\right)  ^{2}}.
\end{align}
Hereafter, we suppose the system can coherently evolve for a sufficiently long
time, i.e., $T\left\vert \omega_{j}-\omega_{j-1}\right\vert \gg1$, where $T$
should not exceed the system decoherence time $T_{\text{coh}}$. Via seeking
the peaks of $F\left(  \omega\right)  $, we can find all the eigen frequencies
$\left\{  \omega_{j}\right\}  _{j=1}^{LW}$. This method can also be applied to
the measurement of Hofstadter-butterfly spectrum (see Appendix.%
~%
\ref{Append:Hof_measurement}).

\bigskip The energy eigenstate $\left\vert E_{j}\right\rangle $ can be
restored from the measured data $\left\vert \tilde{\psi}_{nm}\left(
\omega\right)  \right\rangle $. \ From Eq.%
~%
(\ref{eq:psi_tilde_w}), we find that $\left\vert \tilde{\psi}_{nm}\left(
\omega_{j}\right)  \right\rangle \approx\psi_{nm}^{\left(  j\right)  \ast
}\left\vert E_{j}\right\rangle /\sqrt{2}$, having assumed $T\left\vert
\omega_{j}-\omega_{j-1}\right\vert \gg1$. To remove the ambiguity of the phase
factor of $\left\vert E_{j}\right\rangle $, we implicitly think $\arg\left\{
\psi_{n_{j}m_{j}}^{\left(  j\right)  }\right\}  =0$, which further means
$\arg\left\{  \psi_{nm}^{\left(  j\right)  }\right\}  =-\arg\left\{
\left\langle n_{j}m_{j}\right.  \left\vert \tilde{\psi}_{nm}\left(  \omega
_{j}\right)  \right\rangle \right\}  $. Then, we are convinced to define the
fictitious energy eigenstate $\left\vert \tilde{E}_{j}\right\rangle $, i.e.,
\begin{equation}
\left\vert \tilde{E}_{j}\right\rangle =\sum_{n=-N}^{N}\sum_{m=-M}^{M}\tilde
{M}_{nm}\left(  \omega_{j}\right)  e^{-i\Theta_{nm}\left(  \omega_{j}\right)
}\left\vert \tilde{\psi}_{nm}\left(  \omega_{j}\right)  \right\rangle ,
\end{equation}
where $\Theta_{nm}\left(  \omega_{j}\right)  =\arg\left\{  \left\langle
n_{j}m_{j}\right.  \left\vert \tilde{\psi}_{nm}\left(  \omega_{j}\right)
\right\rangle \right\}  $, and $\tilde{M}_{nm}\left(  \omega_{j}\right)
=2\sqrt{\left\langle \tilde{\psi}_{nm}\left(  \omega_{j}\right)  \right.
\left\vert \tilde{\psi}_{nm}\left(  \omega_{j}\right)  \right\rangle }$. One
can easily prove that in the limit $T\left\vert \omega_{j}-\omega
_{j-1}\right\vert \gg1$, $\tilde{M}\left(  \omega_{j}\right)  \approx\sqrt
{2}\left\vert \psi_{nm}^{\left(  j\right)  }\right\vert $ and $\left\vert
\tilde{E}_{j}\right\rangle \approx\sum_{nm}\left\vert \psi_{nm}^{\left(
j\right)  }\right\vert $ $^{2}\left\vert E_{j}\right\rangle =\left\vert
E_{j}\right\rangle $.
\color{black}%

Having obtained the energy eigenstate $\left\vert E_{j}\right\rangle $ (or
rather, $\psi_{nm}^{\left(  j\right)  }$),\ we also define the quantity
$\CV P^{(j)}\CIV\left(  k_{x}\right)  $ as in\ Eq.%
~%
(\ref{eq:PIj}). Through finding the peaks of $\CV P^{(j)}\CIV\left(  k_{x}\right)  $,
we can establish the dependence between the $\omega$ and $k_{x}^{\prime}$ with
$\hbar\omega\left(  k_{x}^{\prime}\right)  =E_{j}$.

In fact, after the eigenstates and the eigen energies, the Hamiltonian can be
constructed as%
\begin{equation}
\tilde{H}=\sum_{j}E_{j}\left\vert \tilde{E}_{j}\right\rangle \left\langle
\tilde{E}_{j}\right\vert .
\end{equation}
This means that our method offers a way to construct the Hamiltonian of an
unknown system. After the Hamiltonian is reconstructed, or rather in the
tight-binding form, we can also thoroughly analyze the energy bands of the system.

\section{Conclusions}

\label{sec:DC}

We have proposed to engineer an ideal system of the single-particle Harper
Hamiltonian in a two-dimensional architecture based on interacting transmons
mediated by inductive couplers. Through designing a gradiometer form of the
mutual inductance between the coupler and the transmon, the decoherent effect
of the environment noise is believed to be mitigated in some extent. Based on
this architecture, the chiral or topological phenomena induced by effective
magnetic flux are exhaustively studied for the single-particle states
according to the explicit size of the Harper model.

First, we have concentrated on \CV three-leg ladder \CIV
model, which is known as the simplest true-two-dimensional configuration. In
contrast to the quasi-two-dimensional two-leg ladder, the \CV three-leg ladder \CIV
possesses a central row, thus presenting the concept of bulk. We find the
particle current along the central row is always zero, which is fundamentally
due to the equivalence between the space symmetry and the time-reversal
symmetry. \CV Because open boundary three-leg ladder has a finite size, \CIV the interplay between the effective magnetic flux and the
coupling ratio (i.e., the ratio between the column and row coupling strengths) can 
result in the \CV``staggered vortex-Meissner phase transition''\CIV. For the given coupling ratio,
there exists the critical effective magnetic flux, below which, the \CV ``Meissner
phase'' \CIV is maintained. However, if the critical value is exceeded by the
magnitude of the effective magnetic flux, there are staggered transitions
between \CV``vortex phase'' and ''Meissner phase''\CIV, a quite different phenomenon from the
case in the two-leg ladder where only the \CV``vortex phase'' \CIV exists after the
critical effective magnetic flux. In the \CV ``Meissner phase''\CIV, the particle currents
only populates on the edges, which can thus be treated as the analogue to the
quantum Hall effect. The chiral current, defined by the mean particle current
between the top and bottom row resembles the squeezed sinusoidal function of
the effective magnetic flux. When the coupling ratio becomes larger, the
squeezing effect is alleviated and the trend of the curve approaches the
sinusoidal function better. \CV Here, the term ``staggered vortex-Meissner phase'' is only used to describe the staggered transition between the single vortex, denoted by ``Meissner phase'' and multiple vortices, denoted by ``vortex phase'', in a similar manner as in the superconductor
material and two-leg ladder~\cite{Zhao2020PRA,Atala2014NP}. However, it can be verified that no actual phase transition has occurred according to Landau's phase-transition theory.\CIV

Besides, we have  continued to study the multiple-row case of the Harper model.
In detail, we increase the row number beyond three and find the energy
spectrum in the periodical condition gradually approach the topological energy
band. The winding number of the edge states are consistent with the Chern
number of the bulk states. We have estimated the lattice size that can exhibit
the topological energy band. If we only apply the open-boundary condition to
the row direction, the typical Hofstadter-butterfly spectrum will occur. If we
apply the open-boundary condition to both the row and column directions, the
Hofstadter-butterfly spectrum becomes smoother. The Hofstadter-butterfly
spectrum can be measured via investigating the evolution of the single-qubit
raising operator.

We have \CV also \CIV presented two methods on how to measure the topological band structure
in two-dimensional superconducting qubit circuits. The first one is based on
the excitation and measurement of the single-particle eigen states, which
needs the details of the Hamiltonian but requires a loose decoherence time
only if it is enough to accomplish the single-particle state generation. In
contrast, the second one is based on subsequently excitation of all the qubits
and instant measurement of the states during the coherent evolution of a long
enough period, which thus should be guaranteed by a decoherence time long
enough to discern the discrete energy levels. Both methods need the
space-domain Fourier transformation of the wave functions, while, the second
one also needs the Fourier transformation of the wavefunction in the time
domain. Different from the method in Ref.~\cite{Roushan2017Science}, which
mainly focuses on measuring the eigenenergies, our methods have also
systematically discussed how to measure and analyze the energy eigenstates.
The methods proposed here can be generalized to versatile quantum simulation
experiments and are promising for Hamiltonian reconstruction of an unknown system.

\section{Acknowledgments}

Y.J.Z. is supported by Beijing Natural Science Foundation (BNSF) under grants
No.%
~%
4222064, National Natural Science Foundation of China (NSFC) under the grant
No.%
~%
11904013, and State Scholarship Fund.

\appendix

\section{Tunable linear inductive network}

\label{Append:Int-coupler}

\begin{figure}[ptb]
\centering\includegraphics[
width=0.48\textwidth,clip
]{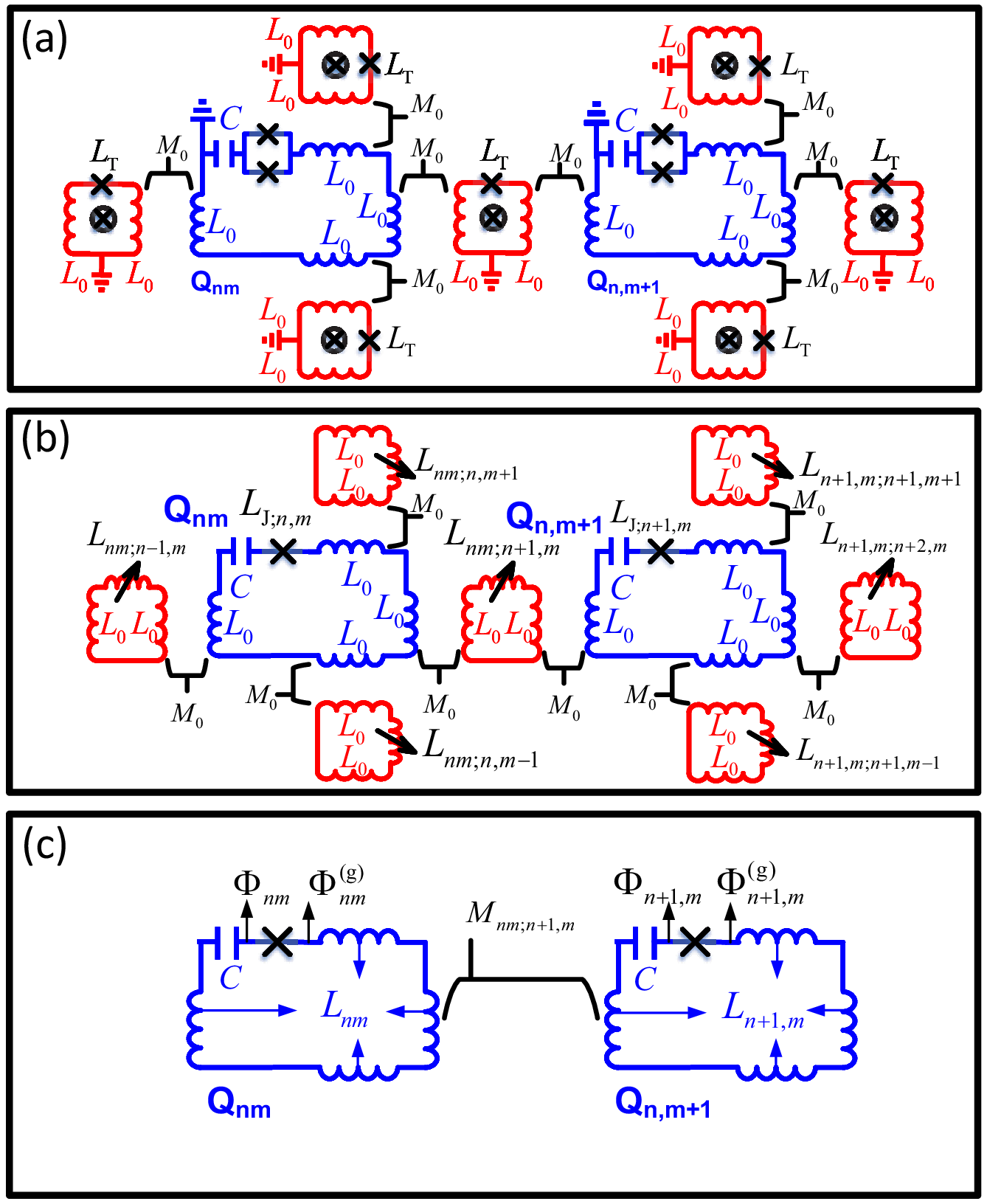}\caption{(a) Coupling circuit diagram between two adjacent qubits
Q$_{nm}$ and\ Q$_{n+1,m}$. The coupler (colored red) and qubit (colored blue)
loop respectively contains two and four segments of self inductance $L_{0}$.
The coupler junction inductance is $L_{\text{T}}$ and the shunting capacitance
of the qubit is $C$. Besides, the mutual inductance between the qubit and
coupler loop is $M_{0}$. (b) Simplified coupling circuit diagram with the
coupler junction treated as a tunable linear inductance ($L_{nm;n^{\prime
}m^{\prime}}$ for the coupler between $nm$ and $n^{\prime}m^{\prime}$) and the
SQUID (superconducting quantum interference device) is replaced by a single
junction (with junction inductance $L_{\text{J;}nm}$ at $nm$). (c) The freedom
of the coupler has been eliminated. This induces a total self inductance
$L_{nm}$ ($L_{n+1,m}$) along the qubit loop at $nm$ ($n+1,m$). The
coupler-mediated interaction is modelled by a mutual inductance $M_{nm;n+1,m}%
$. The node fluxes that construct the Hamiltonian are also indicated (e.g.,
$\Phi_{nm}$, $\Phi_{nm}^{\left(  \text{g}\right)  }$, etc). }%
\label{fig:TLIN}%
\end{figure}

Here, we will show that the coupler and surrounding circuit [see Fig.%
~%
\ref{fig:TLIN}(a)] can be equivalent to a simple tunable linear inductive
network. In detail, we focus on the interaction is between a pair of adjacent
qubits at the site $nm$ and $n^{\prime}m^{\prime}$.

First, we assume zero SQUID branch currents at both sites, that is,
$I_{nm}=I_{n^{\prime}m^{\prime}}=0$. Then, if we assume the flux drop across
the coupler junction is $\Phi^{\text{T}}_{nm;n^{\prime}m^{\prime}}$, the
should be
\begin{equation}
\Phi_{nm;n^{\prime}m^{\prime}}^{\text{T}}+2L_{0}I_{nm;n^{\prime}m^{\prime}%
}-\Phi_{nm;n^{\prime}m^{\prime}}=0\label{eq:LoopFlux}%
\end{equation}
due to trapped magnetic flux in the coupler loop, where $I_{nm;n^{\prime
}m^{\prime}}=I_{\text{c}}\sin\left(  \frac{2\pi\Phi_{nm;n^{\prime}m^{\prime}%
}^{\text{T}}}{\Phi_{0}}\right)  $ is the coupler junction current,
$I_{\text{c}}$ the junction critical current, $\Phi_{0}$ the magnetic flux
quantum, and $L_{0}$ the self inductance of half the coupler loop. Compared to
the junction inductance, which is $L_{\text{T}}=\frac{\Phi_{0}}{2\pi
I_{\text{c}}}$, $L_{0}$ is usually negligible. Thus, we disregard the term
$2L_{0}I_{nm;n^{\prime}m^{\prime}}$ in Eq.%
~%
(\ref{eq:LoopFlux}), yielding $\Phi_{nm;n^{\prime}m^{\prime}}^{\text{T}%
}\approx\Phi_{nm;n^{\prime}m^{\prime}}$.

Now we consider nonzero SQUID branch currents $I_{nm}$ and $I_{n^{\prime
}m^{\prime}}$, which are treated as quantum perturbations to the working point
established by Eq.%
~%
(\ref{eq:LoopFlux}). In this case, the coupler junction is equivalent to a
tunable linear inductor%
~%
\cite{Geller2015PRA} in the quantum regime, with the inductance%
\begin{equation}
L_{nm;n^{\prime}m^{\prime}}=\frac{\Phi_{0}}{2\pi I_{\text{c}}\cos\left(
\frac{2\pi\Phi_{nm;n^{\prime}m^{\prime}}^{\text{T}}}{\Phi_{0}}\right)
}\approx\frac{L_{\text{T}}}{\cos\left(  \frac{2\pi\Phi_{nm;n^{\prime}%
m^{\prime}}}{\Phi_{0}}\right)  }.
\end{equation}
Thus, the circuit in Fig.%
~%
\ref{fig:TLIN}(a) can be simplified into the one in Fig.%
~%
\ref{fig:TLIN}(b) for the perturbative quantum signals.

Last, applying the principle of linear superposition to Fig.%
~%
\ref{fig:TLIN}(b) where the currents $I_{nm}$ and $I_{n^{\prime}m^{\prime}}$
are regarded as sources, we can obtain the node fluxes between the SQUIDs and
their respective grounding wire as
\begin{align}
\Phi_{nm}^{(\text{g})}  &  =L_{nm}I_{nm}+M_{nm;n^{\prime}m^{\prime}%
}I_{n^{\prime}m^{\prime}},\\
\Phi_{n^{\prime}m^{\prime}}^{(\text{g})}  &  =M_{n^{\prime}m^{\prime}%
;nm}I_{nm}+L_{n^{\prime}m^{\prime}}I_{n^{\prime}m^{\prime}}.
\end{align}
Here, $L_{pq}$ and $M_{pq;p^{\prime}q^{\prime}}$ indicate the self and mutual
inductance,\ respectively, which can be represented as
\begin{align}
L_{pq}  &  =4L_{0}+\sum_{p^{\prime}q^{\prime}\in%
\mathbb{C}
_{pq}}M_{pq;p^{\prime}q^{\prime}},\\
M_{pq;p^{\prime}q^{\prime}}  &  =-\frac{M_{0}^{2}}{2L_{0}+L_{pq;p^{\prime
}q^{\prime}}}\nonumber\\
&  =-\frac{M_{0}^{2}\cos\left(  \frac{2\pi}{\Phi_{0}}\Phi_{pq;p^{\prime
}q^{\prime}}\right)  }{L_{\text{T}}+2L_{0}\cos\left(  \frac{2\pi}{\Phi_{0}%
}\Phi_{pq;p^{\prime}q^{\prime}}\right)  },
\end{align}
Recall that $\mathbb{C}_{pq}$ denotes the set 
Thus, Fig.%
~%
\ref{fig:TLIN}(b) can be further simplified into the tunable linear inductive
network in Fig.%
~%
\ref{fig:TLIN}(c).

\section{Degeneracy property of the single-particle ground state}

\label{Append:Degen}

Here, we make some remarks about the single-particle ground state (denoted by
$\psi_{n,m}^{\left(  1\right)  }$ in the main text) in the case of broken
time-reversal symmetry $\left(  0<\left\vert \gamma\right\vert <\pi\right)  $.
First, we assume the length of the double-ladder is sufficiently large (say,
the length $L\rightarrow\infty$). Then, the single-particle ground state in
the open-boundary condition must correspond to the ground bulk state. From
Fig.%
~%
\ref{fig:ChiralCurrent}, we can find that for $0<\left\vert \gamma\right\vert
<\pi$ and not too small $g_{y}/g_{x}$, the minimum energy is achieved at
$k_{x}=0$, where the single-particle ground state must be nondegenerate. For
sufficiently small $g_{y}/g_{x}$, we can prove using the perturbative method
that the eigen energy at $k_{x}=0$, which is
\begin{equation}
E_{0}^{\prime}\approx-2\hbar g_{x}-\frac{\hbar g_{y}^{2}}{g_{x}}\frac
{1}{1-\cos\gamma},\label{eq:E0p}%
\end{equation}
is always lower than those at $k_{x}=\pm\gamma$, which are both%
\begin{equation}
E_{\gamma}^{\prime}=-2\hbar g_{x}+\frac{\hbar g_{y}^{2}}{2g_{x}}\frac
{1}{1-\cos\gamma}.\label{eq:Erp}%
\end{equation}
Thus, the minimum energy is still achieve at $k_{x}=0$, implying that the
single-particle ground state must be nondegenerate.

\begin{figure}[ptb]
\centering\includegraphics[width=0.48\textwidth]{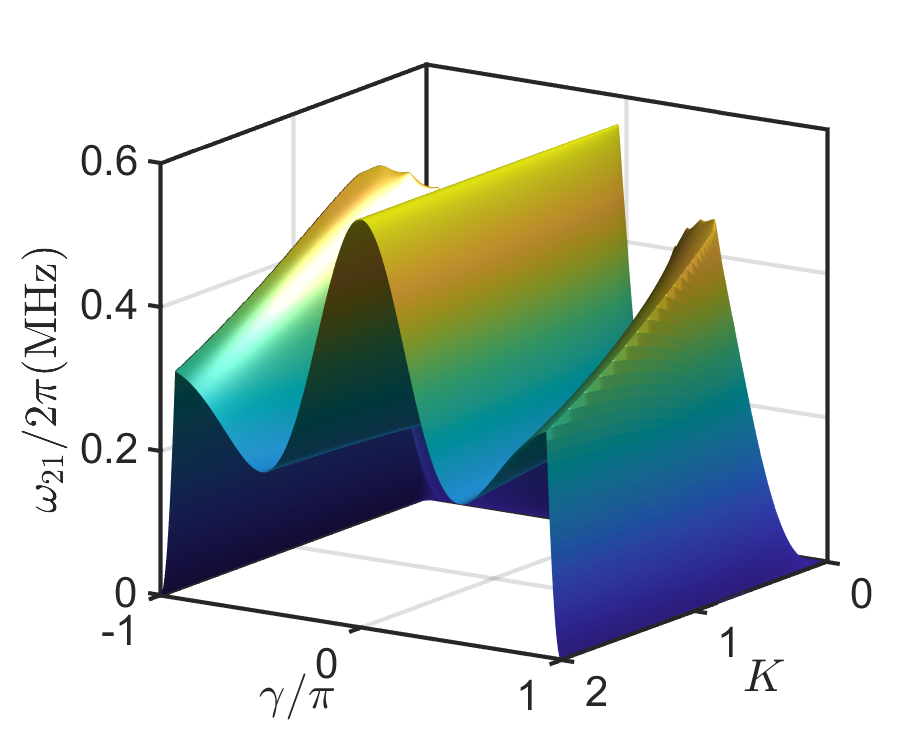}\caption{\color{black}
Frequency interval $\omega_{21}$ between the two lowest energy eigenstates in
the open-boundary condition for the double ladder (i.e., the lattice width
taking $W=3$) plotted against the effective magnetic flux $\gamma$ and
coupling ratio $K$. In this plot, we have specified the lattice length as
$L=17$.\color{black}}%
\label{fig:eigKgamma}%
\end{figure}

However, the have focused on practical lattice length, e.g., $L=17$, which is
far from infinity. The single-particle ground state in the open-boundary
condition may not the ground bulk state. In this case, we have numerically
plotted the frequency interval between the first excited state and the ground
state, i.e., $\omega_{21}=\omega_{2}-\omega_{1}$, against the practical
regimes of $\gamma$ and $K$, just as shown in Fig~\ref{fig:eigKgamma}. We can
conveniently find there is no degeneracy in the regime of interest
$0<\left\vert \gamma\right\vert <\pi$.

In the worst case that the single-particle ground states are degenerate and
$\psi_{n,m}^{\left(  1\right)  \ast}\neq\psi_{n,-m}^{\left(  1\right)  }$, we
recombine the ground states as $\psi_{n,m}^{\left(  1,+\right)  }=\psi
_{n,-m}^{\left(  1\right)  \ast}+\psi_{n,m}^{\left(  1\right)  }$ and
$\psi_{n,m}^{\left(  1,-\right)  \ast}=\psi_{n,-m}^{\left(  1\right)  \ast
}-\psi_{n,m}^{\left(  1\right)  }$, where the normalized constants are
temporarily ignored. One then finds that $\psi_{n,m}^{\left(  1,+\right)
\ast}=\psi_{n,-m}^{\left(  1,+\right)  }$ and $\psi_{n,m}^{\left(  1,-\right)
\ast}=-\psi_{n,-m}^{\left(  1,-\right)  }$, which further induces $\psi
_{n,0}^{\left(  1,+\right)  \ast}=\psi_{n,0}^{\left(  1,+\right)  }$ and
$\psi_{n,0}^{\left(  1,-\right)  \ast}=-\psi_{n,0}^{\left(  1,-\right)  }$.
Thus, both $\psi_{n,m}^{\left(  1,+\right)  }$ and $\psi_{n,m}^{\left(
1,-\right)  }$ heralds a zero particle current along the central row [see Eq.%
~%
(\ref{eq:ChiralCurrent})].

\section{Normalized current pattern}

\label{Append:NormalizedCurrentPattern}

\begin{figure*}[ptbh]
\centering\includegraphics[
width=0.96\textwidth,clip
]{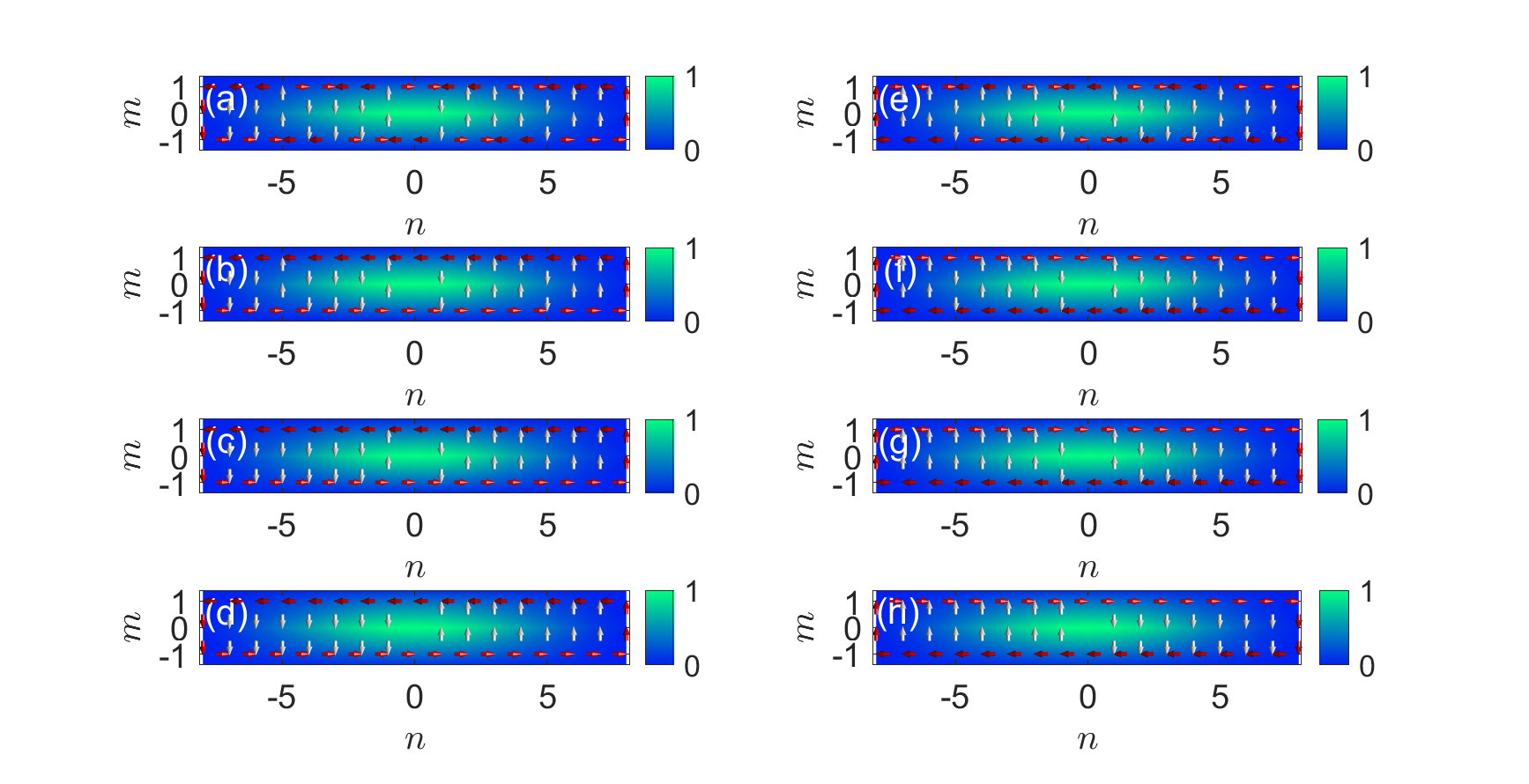}
\caption
{(color online) Normalized particle current patterns for the effective magnetic flux $\gamma
$ taking (a)-(c) $-\frac{\pi}{2}$
and (d)-(f) $\frac{\pi}{2}%
$ in a double ladder (lattice width $W=3$) with  the lattice length length $L=17$. From top to bottom, the coupling strengths along two directions fulfill
the conditions $K=0.1$, $0.2$, $0.4$, and $0.7$, respectively. The particle current between adjacent sites is represented
by a arrow whose size implies the current magnitude.
We have used the red (green) color for currents along the edge (in the bulk).
Beside, the color in the background represents the relative occupation probability in the
single-particle ground state.}
\label{fig:Vortex-Meissner-Normalized}
\end{figure*}

Since the vortex number in Fig.%
~%
\ref{fig:Vortex-Meissner} is hard to discern, we have plotted the current
patterns with all the particle-current amplitudes normalized to one identical
value in Fig.%
~%
\ref{fig:Vortex-Meissner-Normalized}. As has been shown from the top to bottom
row, there are 7, 4, 2, 1 vortices as $K$ takes 0.1, 0.2, 0.4, and 0.7, respectively.%

\section{Calculation of the Chern number\label{Sec:CalChernNum}}

In the one-dimensional chain, the topological property is normally
characterized by Zak phase%
~%
\cite{Bernevig2013Book,Wu2021PRB}, which, however, changes to the Chern number
in the two-dimensional case%
~%
\cite{Bernevig2013Book,Wu2024PRL}. In detail, the Chern number for the $j$th
energy band of our present two-dimensional lattice is defined by
\begin{equation}
C_{j}=\frac{1}{2\pi i}\int\text{d}^{2}\mathbf{k}F_{xy}^{\left(  j\right)
}\left(  \mathbf{k}\right)  ,
\end{equation}
where the integrand is the Berry curvature and is of the form%
\begin{equation}
F_{xy}^{\left(  j\right)  }=\left\langle \partial_{k_{x}}u_{j}\left(
\mathbf{k}\right)  \right.  \left\vert \partial_{k_{y}}u_{j}\left(
\mathbf{k}\right)  \right\rangle
\!%
-%
\!%
\left\langle \partial_{k_{y}}u_{j}\left(  \mathbf{k}\right)  \right\vert
\left.  \partial_{k_{x}}u_{j}\left(  \mathbf{k}\right)  \right\rangle
.\label{eq:Fxy1}%
\end{equation}

Inserting the identity operator $\sum_{r}\left\vert u_{r}\left(
\mathbf{k}\right)  \right\rangle \left\langle u_{r}\left(  \mathbf{k}\right)
\right\vert $ into Eq.%
~%
(\ref{eq:Fxy1}) further yields%
\begin{align}
F_{xy}^{\left(  j\right)  }  &  =\sum_{r}\left\langle \partial_{k_{x}}%
u_{j}\left(  \mathbf{k}\right)  \right.  \left\vert u_{r}\left(
\mathbf{k}\right)  \right\rangle \left\langle u_{r}\left(  \mathbf{k}\right)
\right.  \left\vert \partial_{k_{y}}u_{j}\left(  \mathbf{k}\right)
\right\rangle
\!%
\nonumber\\
&  -%
\!%
\left\langle \partial_{k_{y}}u_{j}\left(  \mathbf{k}\right)  \right\vert
\left.  u_{r}\left(  \mathbf{k}\right)  \right\rangle \left\langle
u_{r}\left(  \mathbf{k}\right)  \right\vert \left.  \partial_{k_{x}}%
u_{j}\left(  \mathbf{k}\right)  \right\rangle \nonumber\\
&  =\sum_{r\neq j}\left\langle \partial_{k_{x}}u_{j}\left(  \mathbf{k}\right)
\right.  \left\vert u_{r}\left(  \mathbf{k}\right)  \right\rangle \left\langle
u_{r}\left(  \mathbf{k}\right)  \right.  \left\vert \partial_{k_{y}}%
u_{j}\left(  \mathbf{k}\right)  \right\rangle
\!%
\nonumber\\
&  -%
\!%
\left\langle \partial_{k_{y}}u_{j}\left(  \mathbf{k}\right)  \right\vert
\left.  u_{r}\left(  \mathbf{k}\right)  \right\rangle \left\langle
u_{r}\left(  \mathbf{k}\right)  \right\vert \left.  \partial_{k_{x}}%
u_{j}\left(  \mathbf{k}\right)  \right\rangle .\label{eq:Fxy2}%
\end{align}
where the terms with $r=j$ can be verified to vanish.

On the other hand, we note that for $r\neq j$, there should be%
\begin{align}
&  \left\langle u_{j}\left(  \mathbf{k}\right)  \right\vert \nabla
_{\mathbf{k}}h\left(  \mathbf{k}\right)  \left\vert u_{r}\left(
\mathbf{k}\right)  \right\rangle \nonumber\\
=  &  \left\langle u_{j}\left(  \mathbf{k}\right)  \right\vert \nabla
_{\mathbf{k}}\left(  h\left(  \mathbf{k}\right)  \left\vert u_{r}\left(
\mathbf{k}\right)  \right\rangle \right)  -\left\langle u_{j}\left(
\mathbf{k}\right)  \right\vert h\left(  \mathbf{k}\right)  \left\vert
\nabla_{\mathbf{k}}u_{r}\left(  \mathbf{k}\right)  \right\rangle \nonumber\\
=  &  \left\langle u_{j}\left(  \mathbf{k}\right)  \right\vert \nabla
_{\mathbf{k}}\left(  E_{r}\left(  \mathbf{k}\right)  \left.  u_{r}\left(
\mathbf{k}\right)  \right\rangle \right)  -E_{j}\left(  \mathbf{k}\right)
\left\langle u_{j}\left(  \mathbf{k}\right)  \right\vert \left.
\nabla_{\mathbf{k}}u_{r}\left(  \mathbf{k}\right)  \right\rangle \nonumber\\
=  &  \nabla_{\mathbf{k}}E_{r}\left(  \mathbf{k}\right)  \left\langle
u_{j}\left(  \mathbf{k}\right)  \right\vert \left.  u_{r}\left(
\mathbf{k}\right)  \right\rangle +E_{r}\left(  \mathbf{k}\right)  \left\langle
u_{j}\left(  \mathbf{k}\right)  \right\vert \left.  \nabla_{\mathbf{k}}%
u_{r}\left(  \mathbf{k}\right)  \right\rangle \nonumber\\
&  -E_{j}\left(  \mathbf{k}\right)  \left\langle u_{j}\left(  \mathbf{k}%
\right)  \right\vert \left.  \nabla_{\mathbf{k}}u_{r}\left(  \mathbf{k}%
\right)  \right\rangle \nonumber\\
=  &  \left(  E_{r}\left(  \mathbf{k}\right)  -E_{j}\left(  \mathbf{k}\right)
\right)  \left\langle u_{j}\left(  \mathbf{k}\right)  \right.  \left\vert
\nabla_{\mathbf{k}}u_{r}\left(  \mathbf{k}\right)  \right\rangle ,
\end{align}
where the gradient operator $\nabla_{\mathbf{k}}=\mathbf{e}_{x}\partial
_{k_{x}}+\mathbf{e}_{y}\partial_{k_{y}}$, the orthogonal relation
$\left\langle u_{j}\left(  \mathbf{k}\right)  \right\vert \left.  u_{r}\left(
\mathbf{k}\right)  \right\rangle =0$ has been used and the result
\begin{equation}
\left\langle u_{j}\left(  \mathbf{k}\right)  \right.  \left\vert
\nabla_{\mathbf{k}}u_{r}\left(  \mathbf{k}\right)  \right\rangle
=\frac{\left\langle u_{j}\left(  \mathbf{k}\right)  \right\vert \nabla
_{\mathbf{k}}h\left(  \mathbf{k}\right)  \left\vert u_{r}\left(
\mathbf{k}\right)  \right\rangle }{E_{r}\left(  \mathbf{k}\right)
-E_{j}\left(  \mathbf{k}\right)  }\label{eq:ukDuk}%
\end{equation}
can hence be obtained.\ 

Having obtained Eq.%
~%
(\ref{eq:ukDuk}), Eq.%
~%
(\ref{eq:Fxy2}) can then be transformed into
\begin{align}
F_{xy}^{\left(  j\right)  }=  &  \left\langle \partial_{k_{x}}u_{j}\left(
\mathbf{k}\right)  \right.  \left\vert \partial_{k_{y}}u_{j}\left(
\mathbf{k}\right)  \right\rangle
\!%
-%
\!%
\left\langle \partial_{k_{y}}u_{j}\left(  \mathbf{k}\right)  \right\vert
\left.  \partial_{k_{x}}u_{j}\left(  \mathbf{k}\right)  \right\rangle
\nonumber\\
=  &  \sum_{r\neq j}\frac{\left\langle u_{j}\right\vert \partial_{k_{x}%
}h\left(  \mathbf{k}\right)  \left\vert u_{r}\right\rangle
\!%
\left\langle u_{r}\right\vert \partial_{k_{y}}h\left(  \mathbf{k}\right)
\left\vert u_{j}\right\rangle }{\left[  E_{j}\left(  \mathbf{k}\right)
-E_{r}\left(  \mathbf{k}\right)  \right]  ^{2}}\nonumber\\
&  -\frac{\left\langle u_{j}\right\vert \partial_{k_{y}}h\left(
\mathbf{k}\right)  \left\vert u_{r}\right\rangle \left\langle u_{r}\right\vert
\partial_{k_{x}}h\left(  \mathbf{k}\right)  \left\vert u_{j}\right\rangle
}{\left[  E_{j}\left(  \mathbf{k}\right)  -E_{r}\left(  \mathbf{k}\right)
\right]  ^{2}}.\label{eq:Fxy3}%
\end{align}
We can see from Eq.%
~%
(\ref{eq:Fxy3}) that a peak or dip can occur where the $j$th band
$E_{j}\left(  \mathbf{k}\right)  $ greatly approaches its adjacent band
$E_{j\pm1}\left(  \mathbf{k}\right)  $.

\section{Hofstadter-butterfly spectrum measurement}

\label{Append:Hof_measurement}

Here, we give details on how to calculate $\chi_{nm}\left(  t\right)
=2\left\langle \sigma_{nm}^{+}\left(  t\right)  \right\rangle $, which
represents the evolution of the single-qubit raising operator. In Sec.%
~%
\ref{Sec:CurrentPattern}, we have introduced single-particle eigenstate
creation operator [see Eq.%
~%
(\ref{eq:Sigma_j_p})], which is
\begin{equation}
S_{j}^{+}=\left\vert E_{j}\right\rangle \left\langle 0\right\vert =\sum
_{n=-N}^{N}\sum_{m=-M}^{M}\psi_{nm}^{\left(  j\right)  }\sigma_{nm}^{+}%
\end{equation}
considering $\sigma_{nm}^{+}\equiv a_{nm}^{\dag}$ in the single-particle
regime. The completeness relation $\sum_{j}\psi_{nm}^{\left(  j\right)  }%
\psi_{n^{\prime}m^{\prime}}^{\left(  j\right)  \ast}=\delta_{nm,n^{\prime
}m^{\prime}}$ further gives
\begin{equation}
\sigma_{nm}^{+}=\sum_{j=1}^{LW}\psi_{nm}^{\left(  j\right)  \ast}S_{j}^{+}.
\end{equation}
Subject to the Harper Hamiltonian [see Eq.%
~%
(\ref{eq:H_sp})], the single-qubit raising operator after time $t$ becomes
\begin{align}
\sigma_{nm}^{+}\left(  t\right)   &  =\sum_{j=1}^{LW}\psi_{nm}^{\left(
j\right)  \ast}S_{j}^{+}e^{i\frac{E_{j}}{\hbar}t}\nonumber\\
&  =\sum_{j=1}^{LW}\psi_{nm}^{\left(  j\right)  \ast}\sum_{n^{\prime}=-N}%
^{N}\sum_{m^{\prime}=-M}^{M}\psi_{n^{\prime}m^{\prime}}^{\left(  j\right)
}\sigma_{n^{\prime}m^{\prime}}^{+}e^{i\frac{E_{j}}{\hbar}t}\nonumber\\
&  =\sum_{n^{\prime}=-N}^{N}\sum_{m^{\prime}=-M}^{M}\sum_{j=1}^{LW}\psi
_{nm}^{\left(  j\right)  \ast}\psi_{n^{\prime}m^{\prime}}^{\left(  j\right)
}\sigma_{n^{\prime}m^{\prime}}^{+}e^{i\frac{E_{j}}{\hbar}t}.
\end{align}
For the initial state $\left(  \left\vert 0\right\rangle +\left\vert
1_{nm}\right\rangle \right)  /\sqrt{2}$, the time evolution of the raising
operator $\sigma_{nm}^{+}$ can be quantified by
\begin{equation}
\chi_{nm}\left(  t\right)  =2\left\langle \sigma_{nm}^{+}\left(  t\right)
\right\rangle =\sum_{j=1}^{LW}\left\vert \psi_{nm}^{\left(  j\right)
}\right\vert ^{2}e^{i\frac{E_{j}}{\hbar}t}\text{.}%
\end{equation}
When averaging $\chi_{nm}$ over all the sites, we can then obtain
\begin{align}
\bar{\chi}\left(  t\right)   &  =\frac{1}{LW}\sum_{n=-N}^{N}\sum_{m=-M}%
^{M}\chi_{nm}\left(  t\right) \nonumber\\
&  =\frac{1}{LW}\sum_{n=-N}^{N}\sum_{m=-M}^{M}\sum_{j=1}^{LW}\left\vert
\psi_{nm}^{\left(  j\right)  }\right\vert ^{2}e^{i\frac{E_{j}}{\hbar}%
t}\nonumber\\
&  =\frac{1}{LW}\sum_{j=1}^{LW}\left(  \sum_{n=-N}^{N}\sum_{m=-M}%
^{M}\left\vert \psi_{nm}^{\left(  j\right)  }\right\vert ^{2}\right)
e^{i\frac{E_{j}}{\hbar}t}\nonumber\\
&  =\frac{1}{LW}\sum_{j=1}^{LW}e^{i\frac{E_{j}}{\hbar}t},
\end{align}
from which, we can see the frequencies that occur in the Fourier
transformation of $\chi\left(  t\right)  $, with a homogeneous amplitude
$\frac{1}{LW}$, are the eigen energies of the Harper Hamiltonian.

Besides, we point out that the Hofstadter-butterfly spectrum can also be
constructed via measuring the time-dependent wave function through
subsequently exciting all the qubits (see Sec.%
~%
\ref{Append:GMMEB}).

\end{document}